\documentclass[fleqn,usenatbib,usedcolumn]{mnras}

\usepackage[british]{babel}             
\usepackage[slantedGreek]{newtxmath}    
\let\la=\lesssim 
\usepackage[T1]{fontenc}                
\usepackage{graphicx}                   

\usepackage{multirow}
\usepackage{subfig}

\newcommand{\btxt}[1]{{\bf #1}}

\title[The Weigelt complex]{On the changes in the physical properties of the\\ ionized region around the Weigelt structures\\ in $\eta$~Carinae over the 5.54-yr spectroscopic cycle\thanks{Based on observations made with the NASA/ESA Hubble Space Telescope, obtained at the Space Telescope Science Institute, which is operated by the Association of Universities for Research in Astronomy, Inc., under NASA contract NAS 5-26555. These observations are associated with program numbers 11506, 12013, 12508, and 12750. Support for program numbers 12013, 12508, and 12750 was provided by NASA through grants from the Space Telescope Science Institute, which is operated by the Association of Universities for Research in Astronomy, Inc., under NASA contract NAS 5-26555.}}

\author[Teodoro et al.]{M.~Teodoro$^{1}$\thanks{CNPq/Science without Borders Fellow}, T.~R.~Gull$^{1,2}$, M.~A.~Bautista$^{3}$, D.~J.~Hillier$^{4,5}$, G.~Weigelt$^{6}$, M.~F.~Corcoran$^{1,7}$\\
$^{1}$Astrophysics Science Division, Code 660, NASA Goddard Space Flight Center, Greenbelt, MD 20771, USA\\
$^{2}$Space Telescope Science Institute, Baltimore, MD 21218,USA\\
$^{3}$Department of Physics, Western Michigan University, Kalamazoo, MI 49008, USA\\
$^{4}$Department of Physics and Astronomy, University of Pittsburgh, 3941 O'Hara Street, Pittsburgh, PA 15260, USA\\
$^{5}$Pittsburgh Particle Physics, Astrophysics and Cosmology Center, University of Pittsburgh, 3941 O'Hara Street, Pittsburgh, PA 15260, USA\\
$^{6}$Max-Planck-Institut f\"ur Radioastronomie, Auf dem H\"ugel 69, D-53121 Bonn, Germany\\
$^{7}${The Catholic University of America, 620 Michigan Ave., N.E. Washington, DC 20064}}

\begin{document}

\date{Final version (\today)}

\pagerange{\pageref{firstpage}--\pageref{lastpage}} \pubyear{2014}

\maketitle

\label{firstpage}

\begin{abstract}
We present \textit{HST/STIS} observations and analysis of two prominent nebular structures around the central source of $\eta$~Carinae, the knots C and D. The former is brighter than the latter for emission lines from intermediate or high ionization potential ions. The brightness of lines from intermediate and high ionization potential ions significantly decreases at phases around periastron. We do not see conspicuous changes in the brightness of lines from low ionization potential ($<$13.6 eV) ions over the orbital period. Line ratios suggest that the total extinction towards the Weigelt structures is $A_V=2.0$. Weigelt C and D are characterized by an electron density of $10^{6.9}$~cm$^{-3}$ that does not significantly change throughout the orbital cycle. The electron temperature varies from $5500$~K (around periastron) to $7200$~K (around apastron). The relative changes in the brightness of \ion{He}{i} lines are well reproduced by the  variations in the electron temperature alone. We found that, at phases around periastron, the electron temperature seems to be higher for Weigelt C than that of D. The Weigelt structures are located close to the Homunculus equatorial plane, at a distance of about 1240~AU from the central source. From the analysis of proper motion and age, the Weigelt complex can be associated with the equatorial structure called `Butterfly Nebula' surrounding the central binary system.
\end{abstract}


\section{Introduction}
\begin{figure}
  \centering
  \includegraphics[width=\linewidth]{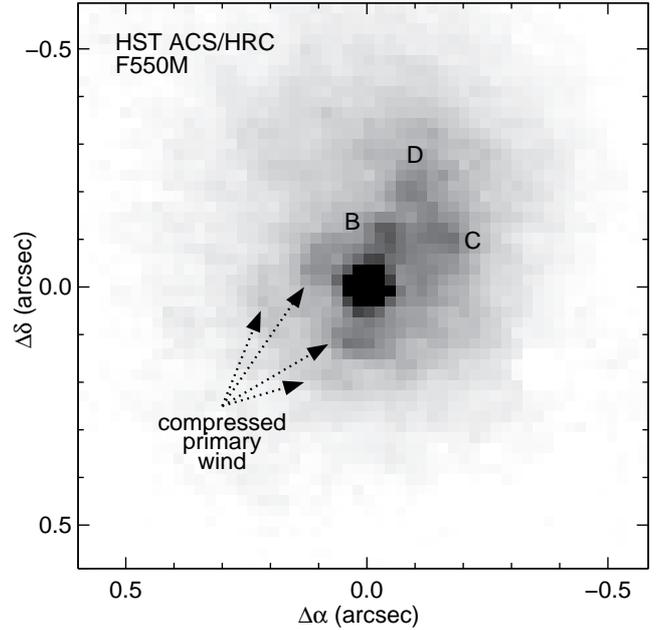}
  \caption{Archival \textit{HST} image obtained on 1992 Nov. 03 with the Advanced Camera for Surveys/High Resolution Camera (\textit{ACS/HRC}). In this image, the optical Weigelt knots B, C, and D can be clearly identified. The compressed primary wind structures \citep[discussed in detail in][]{2013ApJ...773L..16T} are also indicated. North is up and East is to the left.\label{fig:opticalblobs}}
\end{figure}

\btxt{The Weigelt knots\footnote{\btxt{While often called the Weigelt 'blobs', we choose to call them the Weigelt knots.}} were first observed in the optical region through speckle interferometric techniques \citep{1986A&A...163L...5W}. These structures are responsible for the conspicuous, low velocity, narrow emission lines typically seen superimposed on the emission lines in the spectrum of the massive, wind-wind colliding binary system in $\eta$~Car \cite[][]{1995AJ....109.1784D}. In the observer's line of sight to the central source, the total extinction is higher than in the line of sight to the knots, and that makes them appear unusually bright when compared to the central source \citep[see Fig.~\ref{fig:opticalblobs} and][]{1992A&A...262..153H,1995AJ....109.1784D}.}

\btxt{Many studies have contributed to determine the nature and to constrain the physical properties and spatial orientation of the Weigelt knots. An extensive and detailed review of the nebular equatorial structures seen in $\eta$~Car was done by \cite{1997ARA&A..35....1D} and \cite{2012ASSL..384..129W}, and we refer the reader to these publications for more information. Below, for the sake of brevity, we mention a few key publications that are relevant to the present paper.}

\btxt{\citet{1995AJ....109.1784D,1997AJ....113..335D} showed that the knots were condensed nebular ejecta and that they lie close to the equatorial region, with an estimated H\,{\sc i} density and electron temperature of $N_{\rmn{H\,I}}>10^7$~cm$^{-3}$ and $T_{\rmn e}\approx8000$~K, respectively. \cite{1999ASPC..179..116H} showed that the knots present a partially-ionized region responsible for the weak H\,{\sc i} and strong Fe\,{\sc ii} emission, and that they are CNO-processed material, with overabundance of N and depletion of C and O, relative to solar values \cite[see also][]{2005ApJ...624..973V}. Those authors also concluded that the electron density of the knots should be within the \mbox{$N_{\rmn e}\sim10^{4}$--$10^{8}$~cm$^{-3}$} range.}

\btxt{\cite{2001A&A...378..266J,2005MNRAS.364..731J} showed that selective photo-excitation by resonant H\,{\sc i} transitions are required in order to explain the anomalously high intensity of the Fe\,{\sc ii}~$\lambda2507$ and $\lambda2509$ lines, as well as the behavior of the O\,{\sc i} lines at $8446$ (fluorescent), $6300$ (forbidden transition), and $7774$~\AA (allowed transition). Additionally, \cite{2002ApJ...581.1154V,2005ApJ...624..973V} showed that continuum pumping is also important to appropriately reproduce the intensity of the optical Fe II lines.}

\begin{figure}
  \centering
  \includegraphics[width=\linewidth]{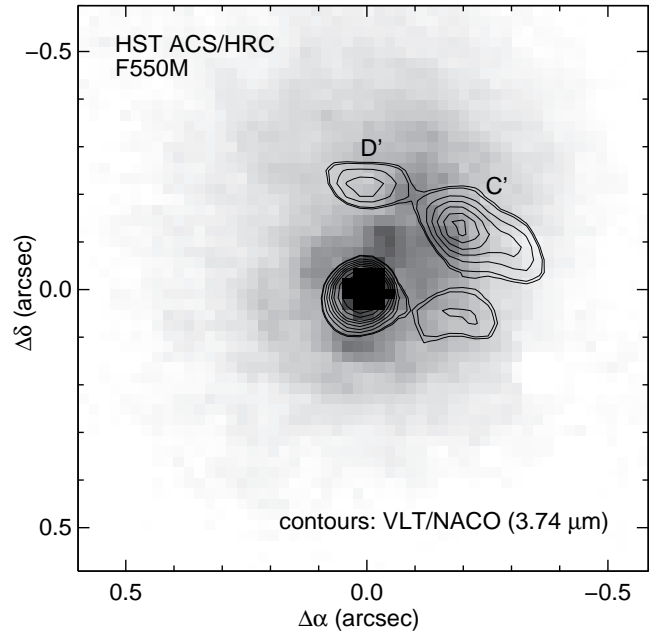}
  \caption{Comparison between the optical knots and the infrared structures detected on NACO images obtained on 2002 Nov 16 by \protect\cite{2005A&A...435.1043C} using the NB\_374 filter (central wavelength around 3.74~$\mu$m; dominated by emission from the hydrogen line Pf$\gamma$). The position of the NACO structures was corrected for the difference of 10~yr between the images.\label{fig:opticalblobs2}}
\end{figure}

\begin{table*}
    \centering
    \caption{Summary of the HST/STIS observations of $\eta$~Carinae.}\label{tab:obslog}
    \begin{tabular}{ccccccccc}

    \hline\hline
    \multirow{2}{*}{Program} &
    \multirow{3}{*}{P.\,I.} &
    \multirow{3}{*}{Grating} &
    Central &
    Mapping &
    Pixel &
    \multirow{2}{*}{Slit} &
    \multirow{2}{*}{Observation} &
    \multirow{3}{*}{$\Phi$\textsuperscript{a}}
    \\
    \multirow{2}{*}{ID} &
     &
     &
    wavelength &
    region size &
    scale &
    \multirow{2}{*}{PA} &
    \multirow{2}{*}{date} &
    \\
     &
     &
     &
    (\AA) &
    (arcsec$^2$) &
    (arcsec pixel$^{-1}$) &
     &
     & \\\hline
     \multirow{2}{*}{11506} & \multirow{2}{*}{K. Noll} & G430M & 4706 & \multirow{2}{*}{6.4$\times$2.0} & \multirow{2}{*}{0.10} & \multirow{2}{*}{$+79.51\degr$} & \multirow{2}{*}{2009 Jun 30} & \multirow{2}{*}{12.085} \\
 & & G750M & 5734, 7283 & & & & & \\ \cline{3-4}
\multirow{2}{*}{12013} & \multirow{2}{*}{M. Corcoran} & G430M & 4706 & \multirow{2}{*}{6.4$\times$1.4} & \multirow{2}{*}{0.10} & \multirow{2}{*}{$-121.03\degr$} & \multirow{2}{*}{2009 Dec 06} & \multirow{2}{*}{12.163} \\
 & & G750M & 5734, 7283 & & & & & \\ \cline{3-4}
\multirow{2}{*}{12013} & \multirow{2}{*}{M. Corcoran} & G430M & 4706 & \multirow{2}{*}{6.4$\times$1.1} & \multirow{2}{*}{0.05} & \multirow{2}{*}{$-166.67\degr$} & \multirow{2}{*}{2010 Oct 26} & \multirow{2}{*}{12.323} \\
 & & G750M & 5734 & & & & & \\ \cline{3-4}
\multirow{2}{*}{12508} & \multirow{2}{*}{T. Gull} & G430M & 4706 & \multirow{2}{*}{6.4$\times$2.0} & \multirow{2}{*}{0.05} & \multirow{2}{*}{$-138.66\degr$} & \multirow{2}{*}{2011 Nov 20} & \multirow{2}{*}{12.516} \\
 & & G750M & 5734, 7283 & & & & & \\ \cline{3-4}
\multirow{2}{*}{12750} & \multirow{2}{*}{T. Gull} & G430M & 4706 & \multirow{2}{*}{6.4$\times$2.0} & \multirow{2}{*}{0.05} & \multirow{2}{*}{$-174.84\degr$} & \multirow{2}{*}{2012 Oct 18} & \multirow{2}{*}{12.680} \\
 & & G750M & 5734, 7283 & & & & & \\ \cline{3-4}
\multirow{2}{*}{13054} & \multirow{2}{*}{T. Gull} & \multirow{2}{*}{G430M} & \multirow{2}{*}{4706} & \multirow{2}{*}{6.4$\times$2.0} & \multirow{2}{*}{0.05} & \multirow{2}{*}{$-136.73\degr$} & \multirow{2}{*}{2013 Sep 03} & \multirow{2}{*}{12.839} \\
 & & &  & & & & & \\\hline\hline
     \multicolumn{9}{l}{\textsuperscript{a}\footnotesize{$\Phi=12+\varphi$, where $\varphi$ is the \btxt{phase of the spectroscopic cycle}. See text for details.}}
    
    \end{tabular}
\end{table*}

\begin{table*}
    \centering
    \caption{List of the spectral lines used in this work.\label{tab:linelist}}
    \begin{tabular}{cccrrrrc}

    \hline\hline
    \multirow{2}{*}{Line ID\textsuperscript{a}} &
    Wavelength\textsuperscript{b} &
    Transition &
    $E_{1}$ &
    $E_{2}$ &
    $\chi_1$\textsuperscript{c} &
    $\chi_2$\textsuperscript{d} &
    $n_\rmn{c}$\textsuperscript{e}
    \\
     &
    (\AA) &
    (lower $-$ upper) &
    (eV) &
    (eV) &
    (eV) &
    (eV) &
    ($10^6$~cm$^{-3}$) \\\hline\hline
   
    ~[Ar III]~$\lambda$7137 & 7137.76 & $^3$P$_{2}$ $-$ $^1$D$_{2}$ & 0.000 & 1.737 & 27.6 & 40.7 & $4.8$\\
    He I~$\lambda$4714 & 4714.48 & $^3$P $-$ $^3$S & 20.964 & 23.593 & 0.0 & 24.6 & N.\,A.\\
    He I~$\lambda$7065 & 7067.18 & $^3$P $-$ $^3$S & 20.964 & 22.718 & " & " & N.\,A.\\
    He I~$\lambda$7283 & 7283.37 & $^1$P$_{1}$ $-$ $^1$S$_{0}$ & 21.217 & 22.920 & " & " & N.\,A.\\
    ~[Fe III]~$\lambda$4608 & 4607.03 & $^5$D$_{4}$ $-$ $^3$F2$_{3}$ & 0.000 & 2.690 & 16.2 & 30.7 & $4.9$ \\
    ~[Fe III]~$\lambda$4659 & 4658.05 & $^5$D$_{4}$ $-$ $^3$F2$_{4}$ & 0.000 & 2.661 & " & " & $7.4$ \\
    ~[Fe III]~$\lambda$4702 & 4701.53 & $^5$D$_{3}$ $-$ $^3$F2$_{3}$ & 0.054 & 2.690 & " & " & $4.9$ \\
    ~[Fe III]~$\lambda$4735 & 4733.91 & $^5$D$_{2}$ $-$ $^3$F2$_{2}$ & 0.092 & 2.710 & " & " & $2.4$ \\
    ~[Fe III]~$\lambda$4756 & 4754.69 & $^5$D$_{3}$ $-$ $^3$F2$_{4}$ & 0.054 & 2.661 & " & " & $7.4$ \\
    ~[Fe III]~$\lambda$4771 & 4769.43 & $^5$D$_{2}$ $-$ $^3$F2$_{3}$ & 0.092 & 2.690 & " & " & $4.9$ \\
    ~[N II]~$\lambda$5756 & 5756.21 & $^1$D$_{2}$ $-$ $^1$S$_{0}$ & 1.899 & 4.053 & 14.5 & 29.6 & $16.4$\\
    ~[Fe II]~$\lambda$4641 & 4640.97 & a6d $^6$D$_{3/2}$ $-$ b4p $^4$P$_{1/2}$ & 0.107 & 2.778 & 7.9 & 16.2 & $3.6$\\
    ~[Fe II]~$\lambda$4730 & 4729.39 & a6d $^6$D$_{5/2}$ $-$ b4p $^4$P$_{3/2}$ & 0.083 & 2.704 & " & " & $6.0$\\
    ~[Fe II]~$\lambda$4776 & 4776.05 & a4f $^4$F$_{9/2}$ $-$ b4f $^4$F$_{7/2}$ & 0.232 & 2.828 & " & " & $20.2$\\
    ~[Fe II]~$\lambda$4815 & 4815.88 & a4f $^4$F$_{9/2}$ $-$ b4f $^4$F$_{9/2}$ & 0.232 & 2.807 & " & " & $45.0$\\
    ~[Ni II]~$\lambda$7413 & 7413.65 & $^2$D$_{3/2}$ $-$ $^2$F$_{5/2}$ & 0.187 & 1.859 & 7.6 & 18.2 & $1.4$\\
    ~[Ni II]~$\lambda$7379 & 7379.86 & $^2$D$_{5/2}$ $-$ $^2$F$_{7/2}$ & 0.000 & 1.680 & " & " & $0.6$\\\hline\hline
    \multicolumn{7}{l}{\textsuperscript{a}\footnotesize{As used throughout the text and in the figures.}} \\
    \multicolumn{7}{l}{\textsuperscript{b}\footnotesize{Rest wavelength in vacuum.}} \\
    \multicolumn{7}{l}{\textsuperscript{c}\footnotesize{Ionization potential from previous ionization stage ($X^{i-1} \rightarrow X^{i}$).}} \\
    \multicolumn{7}{l}{\textsuperscript{d}\footnotesize{Ionization potential to the next ionization stage ($X^{i} \rightarrow X^{i+1}$).}} \\
    \multicolumn{7}{l}{\textsuperscript{e}\footnotesize{Critical density of upper level at $T_\rmn{e}=10^4$~K.}}

    \end{tabular}

\end{table*}

\btxt{A detailed map of the actual shape and distribution of the material located in the equatorial region was made possible by combining high spatial resolution imaging with spectroscopy. \cite{2002ApJ...567L..77S} and \cite{2005A&A...435.1043C} showed the presence of a spatially extended, clumpy, and dusty ring-like structure (called the Butterfly Nebula) that becomes apparent at wavelengths longward than 2~$\mu$m. Without any information on the kinematics of this ring-like structure, \cite{2005A&A...435.1043C} suggested that it could be the projection on the sky of a bipolar nebula, but \cite{2006ApJ...644.1151S} showed that this structure lies close to the equatorial plane between the two bipolar lobes of the Homunculus.}

\btxt{As noted by \cite{2005A&A...435.1043C}, the position of the optical Weigelt knots does not match the position of the infrared bright structures (named C$^\prime$ and D$^\prime$ by those authors; Fig.~\ref{fig:opticalblobs2}). The optical structures are actually emission from the partially-ionized region around the clumps seen in the infrared. They find no infrared counterpart for the optical knot B. In contrast, there is no optical counterpart for the infrared structure located to the West of the central source.}

\btxt{Another remarkable feature of the Weigelt knots is that they respond to changes in the incident radiation. \cite{1998A&AS..133..299D} originally noted the changes in nebular emission line intensities as a function of excitation energy in the spectrum of Eta Carinae. \cite{2005A&A...436..945H}, \cite{2005ASPC..332..283H}, \cite{2005ApJ...624..973V}, \cite{2008MNRAS.386.2330D},  \cite{2010ApJ...710..729M} and \cite{2016MNRAS.462.3196G} showed that, throughout the 5.5-yr spectroscopic cycle, drastic changes in the intensity of emission occur for the high ionization potential ions (e.g. Fe$^{2+}$, He$^{+}$, Ar$^{2+}$, Ne$^{2+}$), whereas the emission from low ionization potential ions (e.g. Fe$^{+}$ and Ni$^{+}$) remain roughly constant. This kind of behavior allowed \cite{2002ApJ...581.1154V}, \cite{2005ApJ...624..973V}, and \cite{2010ApJ...710..729M} to employ photo-ionization models to constrain the physical properties of the knots and, thus, indirectly probe the nature of the secondary star. Although these studies agree that $10^6<N_{\rmn e}\lesssim2\times10^7$~cm$^{-3}$, they obtain appreciably different parameters for the secondary, especially regarding its effective temperature and luminosity.}

\btxt{In this work, we present a detailed analysis of the changes in the physical properties of the Weigelt knots over almost an entire spectroscopic cycle with complete spatial coverage of the inner region \btxt{($1.6\times1.6$~arcsec$^2$)} around central source. This allowed us to determine the physical properties for each one of the Weigelt knots across the spectroscopic cycle, characterizing the changes in the ionization structure, brightness, electron density and temperature, and estimating the distance and orientation of the knots regarding the central source. 
\cite{2016MNRAS.462.3196G} also did three dimensional spectroscopy of the Weigelt knots.} 

\btxt{Previous work only made use of observations with a single slit  that was either aligned along only one pre-defined direction or did not covered the Weigelt complex entirely due to successive random orientations defined by {\it HST}. This work is the first to present full spatial information on the distribution of the flux from low, intermediate, and high ionization potential ions, across the broad spectroscopic cycle.}

\btxt{We adopt the same nomenclature as defined in \cite{2005A&A...435.1043C}, where the optical Weigelt knots are named B, C, and D, whereas the infrared structures are named C$^\prime$ and D$^\prime$. The paper is organized as follows. In Section \ref{sec:obs} we describe the observations, data reduction and analysis, whereas the results are presented in Section \ref{sec:results}. We discuss our results in Section \ref{sec:discussion} and summarize our conclusions in Section \ref{sec:conclusions}.}

\section{Observations, data reduction, and analysis}\label{sec:obs}
Since we present analysis of data obtained at different phases of the orbital cycle, we must indicate the ephemeris equation we adopt throughout this paper. It is given by \mbox{JD=JD$_0$+2022.7$\times$($\Phi - 13$)}, where \mbox{JD$_0=2,456,874.4$} refers to the extrapolated time of minimum intensity of the narrow component of the He\,{\sc i}~$\lambda 6678$ line \citep{2016ApJ...819..131T} and $\Phi$ is the cycle+phase counting. Cycle 12 extends from 2009.0 to 2014.5. Hence, in the present work, $\Phi=12+\varphi$, where $\varphi$ is the phase of the spectroscopic cycle ($0 \leq \varphi <  1$).

A series of spatio-spectral maps around the central region around $\eta$~Car have been obtained by the Hubble Space Telescope (\textit{HST}) using the Space Telescope Imaging Spectrograph (\textit{STIS}). \btxt{It should be noted that the dataset for phases 0.085, 0.163, and 0.323, presented in this work, is the same as in \cite{2011ApJ...743L...3G}}.

Previous \textit{HST/STIS} long slit observations planned and executed during the \textit{HST} Treasury Program on Eta Carinae\footnote{\url{http://etacar.umn.edu}} (\textit{HST} program numbers 9420 and 9973, \btxt{PI K.~Davidson}) focused on measuring the changes of $\eta$~Car and the Weigelt knots B and D at critical phases of the 5.54-yr orbital period. Due to solar panel constraints, \textit{HST} could accommodate only a limited range of slit position angles that changed from visit to visit. As the primary focus was on the spectrum of the central source of $\eta$~Car, critically timed observations occurred at \textit{HST}-defined position angles (PA). Multiple visits from these and previous programs extended from 1998.0 to 2004.3.

The long slit \textit{HST/STIS} observations recorded extended forbidden emission, most notably of Fe$^{+}$ and Fe$^{2+}$, structures that was found to originate in the interacting winds of the binary system \btxt{\citep{2009MNRAS.396.1308G,2011ApJ...743L...3G,2016MNRAS.462.3196G}}. The \textit{HST/STIS} maps reveal dramatic changes of flux and spatial distribution in emission from ions with low (\mbox{$<13.6$~eV}) and high (\mbox{$>13.6$~eV}) ionization potentials\footnote{\btxt{The classification between low and high ionization potential is based on the comparison between the ionization energy of the ion and that of H\,{\sc i}. We use this distinction due to the fact that only the primary's wind can produce significant amounts of photons with energy \mbox{$<13.6$~eV}.}} near periastron passage. Three-dimensional smooth particle hydrodynamic (\textit{SPH}) modeling \citep{2012MNRAS.420.2064M} of the interacting winds replicated the individual samples of forbidden emissions. However, a complete picture of how the interacting winds changed with orbital phase could not be made. Moreover, long term monitoring of the X-ray flux and H$\alpha$ equivalent width over several recent cycles, showed orbit-to-orbit changes in $\eta$~Car. Individual spectral slices recorded in a single 5.54-year orbit could not determine whether the winds of the binary companions were changing.

Complete maps of the spatially varying forbidden emission were needed. However, the \textit{STIS} instrument failed in the summer of 2004. With the \textit{STIS} repair during the \textit{HST} Servicing Mission 4, in 2009 May, operational capability of the instrument was demonstrated by mapping $\eta$~Car as an Early Release Observation (\textit{HST} program number 11506). A $6.4\times2$ arcsec$^2$ field was sampled at 0.1 arcsec spacing (\textit{i.e.} 0.1 arcsec slit shifts) with the $52\times0.1$~arcsec$^2$ slit aperture at three grating settings of the moderate ($R=\lambda/\Delta\lambda=8000$) dispersion gratings (Table~\ref{tab:obslog}) permitting complete spatial maps of forbidden emission lines \btxt{from ions} such as Fe$^{+}$, Ni$^{+}$, N$^{+}$, Fe$^{2+}$, and Ar$^{2+}$.

This initial mapping captured the early re-emergence (at $\varphi=0.085$) of the colliding wind `bow shock' in front of the primary star ($\eta$~Car A), right after the 2009.0 periastron passage of the secondary ($\eta$ Car B) to within one to two AU behind $\eta$ Car A. A second set of maps with the same spatial sampling of 0.1~arcsec was accomplished in December 2009 (at $\varphi=0.163$). Analysis of these maps confirmed that the instrumental response was about 0.11 arcsec full-width at half-maximum (\textit{FWHM}) so that a higher spatial sampling (0.05 arcsec) was used in the third mapping observation in October 2010 ($\varphi=0.323$) and in all of the following observations through September 2013 ($\varphi=0.839$).

The data was reduced with standard {\sc idl}\footnote{{\sc idl} is a trademark of Exelis Visual Information Solutions, Inc.; \url{http://www.exelisvis.com/ProductServices/IDL.aspx}} data tools. Special data analysis tools were developed to resample the multiple spectra into a standard grid referenced to right ascension and declination. Data cubes were extracted in the form of velocity (20 km\,s$^{-1}$ intervals\footnote{\btxt{Note, however, that the spectral resolution of STIS is about 38~km\,s$^{-1}$.}}), right ascension (0.05 arcsec\,pixel$^{-1}$) and declination (0.05 arcsec\,pixel$^{-1}$). As the first two visits have a coarser pixel scale (0.1 arcsec\,pixel$^{-1}$) than those taken subsequently, we resampled the early observations to keep the whole dataset with the same pixel scale of 0.05 arcsec\,pixel$^{-1}$ (the total flux for each dataset was conserved, though). Each data cube included extractions from $-500$ to $+500$ km\,s$^{-1}$. For each position, continuum was subtracted based upon selected continuum points found to be devoid of nebular line and wind line emission.

\btxt{In the present paper, for all line ratio diagnostics used, we measured the total flux of the Weigelt knots by integrating the line emission within a region centered on $\Delta\alpha=-0.25$ arcsec; $\Delta\delta=-0.13$ arcsec for Weigelt C, and $\Delta\alpha=-0.05$ arcsec; $\Delta\delta=-0.38$ arcsec for Weigelt D. Three windows with different dimensions were used to estimate the uncertainties: $0.20\times0.20$ (width in right ascension$\times$height in declination), $0.25\times0.25$, and $0.30\times0.25$ arcsec$^2$. The total flux for each line of the selected line ratio pair was then divided by the corresponding flux-integration area ($0.20\times0.20$, $0.25\times0.25$, or $0.30\times0.30$ \arcsec$^2$), resulting in the specific intensity (or brightness) of each knot.}

\section{Results}\label{sec:results}

\begin{figure*}
  \centering
  \includegraphics[width=\linewidth]{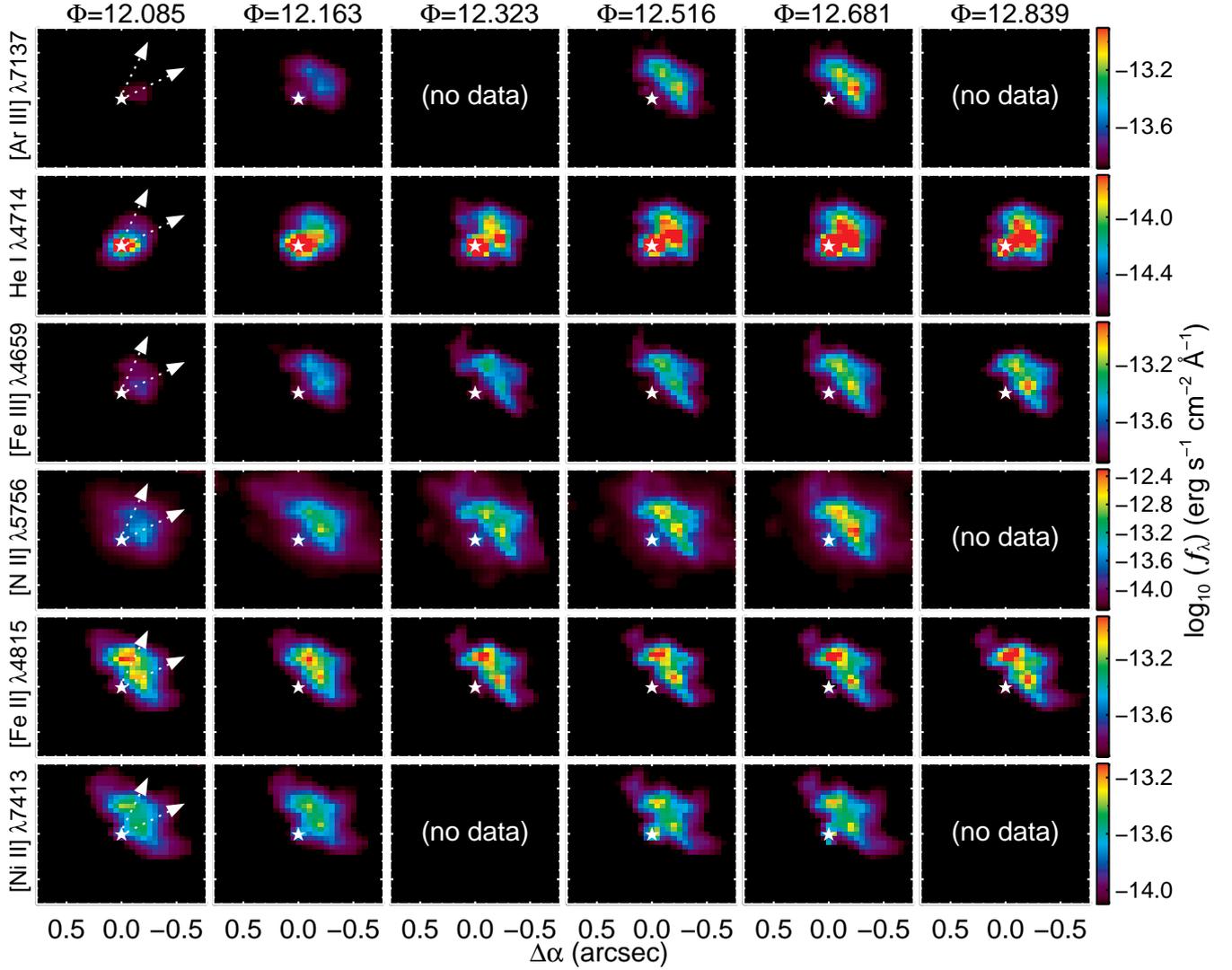}
  \caption{\btxt{Continuum-subtracted} iso-velocity images of the Weigelt complex. The identification of the transition is indicated at the left of each row and phase of the spectroscopic cycle is labelled at the top of each column. The iso-velocity images ($\delta V=20$~km\,s$^{-1}$) show the emission coming from a plasma moving at $-40$~km\,s$^{-1}$, which corresponds to the observed Doppler velocity of the Weigelt complex. The arrows indicate the direction and length of the region we used to extract the intensity profiles shown in Fig.~\ref{fig:profiles1}. \btxt{The color scale for each transition was chosen so as to highlight the emission coming from the two Weigelt knots individually.} North is up and East is to the left.\label{fig:isovelocity}}
\end{figure*}

\begin{figure*}
  \centering
  \includegraphics[width=\linewidth]{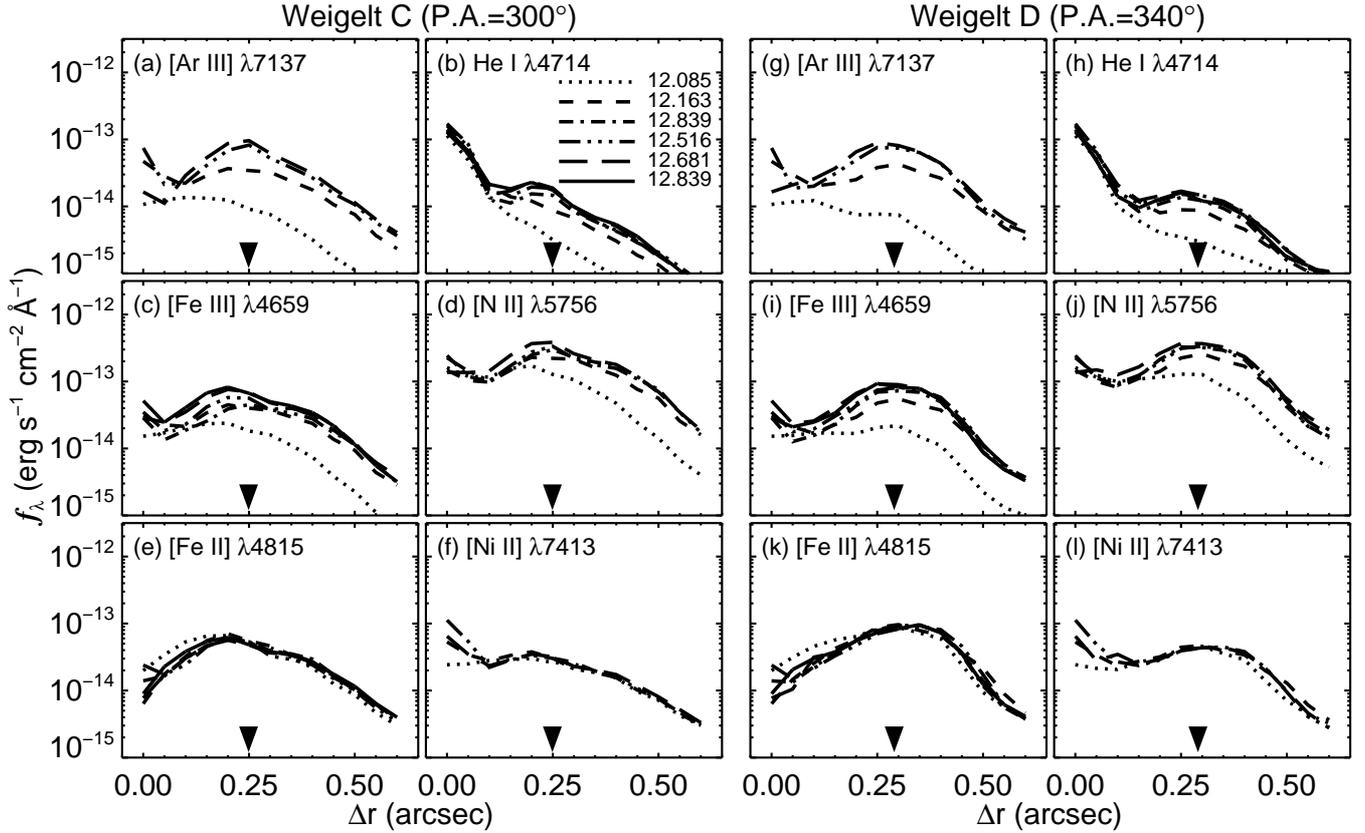}
  \caption{Time variation of the radial flux profiles of Weigelt C (left panel) and D (right panel) for selected representative transitions of high (top row), intermediate (middle row), and low (bottom row) excitation lines. These profiles were extracted exactly along the PA indicated at the top of the two panels (indicated by the arrows in Fig.~\ref{fig:isovelocity}). The triangle indicates the updated position for each knot based on the results from \citet{2004ApJ...605..405S}. Note that the triangle matches the maximum emission for the intermediate/high ionization potential ions and not the low ones (see text for details). Different curves correspond to different phases of the spectroscopic cycle.\label{fig:profiles1}}
\end{figure*}

\begin{figure}
  \centering
  \includegraphics[width=\linewidth]{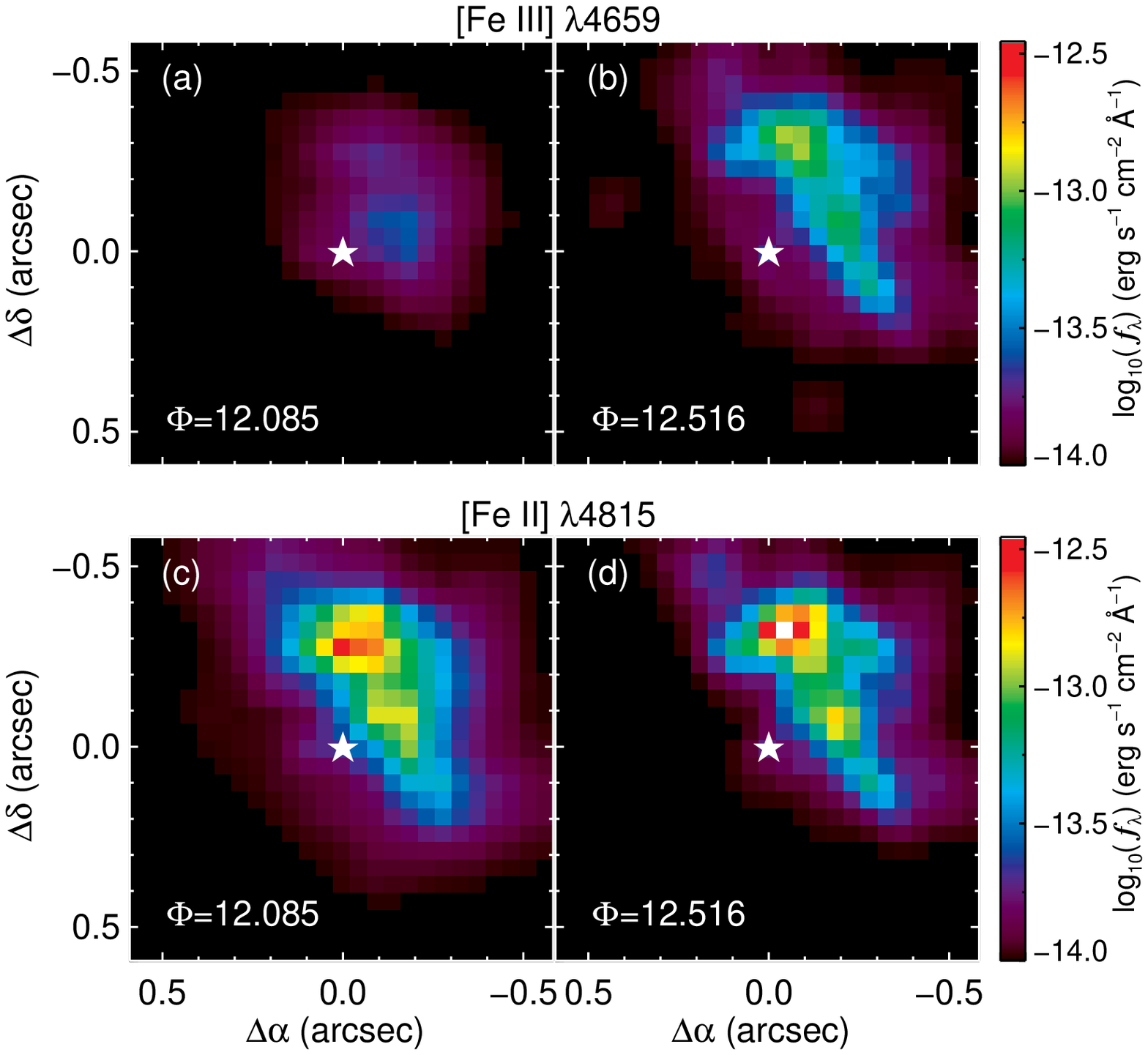}
  \caption{Ionization structure of the Weigelt complex right after periastron (left panels) and around apastron (right panels). From (a) late periastron to (b) apastron, both the flux and the size of the Fe$^{2+}$ emitting region increase considerably. In contrast, the Fe$^{+}$ region decreases in size (c--d), although the overall Fe$^{+}$ flux does not change. North is up and East is to the left.\label{fig:profiles2}}
\end{figure}

\subsection{Changes in the spatial distribution of line emission in the Weigelt complex\label{sec:2D}}

Fig.~\ref{fig:isovelocity} shows the \btxt{spatially-resolved} distribution of emission from a sample of low, intermediate, and high ionization potential ions during almost \btxt{an entire} orbital period\btxt{, while Fig.~\ref{fig:profiles1} shows the intensity profiles extracted from two selected directions (indicated by the arrows in Fig~\ref{fig:isovelocity}) corresponding to the average location of the Weigelt knots as determined by \cite{2004ApJ...605..405S} and \cite{2004AJ....127.1052D}, and corrected by the proper motion determined by those authors.}

\btxt{The first thing to note about the Weigelt knots is that the line emission associated with them is extended and diffuse, as opposed to point-like sources as seen in optical (continuum plus emission lines) images. The spatial distribution, of course, varies with the phase of the spectroscopic cycle and is also dependent on the line transition observed. For instance, the emission from high ionization potential ions (\textit{e.g.} Ar$^{2+}$ and He$^{0}$), as well as from intermediate ones (\textit{e.g.} Fe$^{2+}$ and N$^{+}$), is strongly dependent on the phase of the spectroscopic cycle (see Fig~\ref{fig:profiles1}\textit{a--d} and \textit{g--j}). On the other hand, emission from low ionization ions (Fig~\ref{fig:profiles1}\textit{e}, \textit{f}, \textit{k}, and \textit{l}) remains roughly constant throughout the spectroscopic cycle.}

\btxt{Although the behavior mentioned in the previous paragraph has been reported by previous works \cite[e.g.][]{2005ASPC..332..283H,2005ApJ...624..973V,2010ApJ...710..729M}, Fig.~\ref{fig:isovelocity} shows that not only the flux of the emission line depends on the phase spectroscopic cycle, but the projected shape of the emission region, as well. Indeed, emission from high ionization potential ions is distributed over a more compact area relative to that from low ionization potential ions.}

\btxt{A comparison between Fig.~\ref{fig:isovelocity} and \ref{fig:profiles1} clearly shows that, for intermediate and high ionization potential ions, the increase in the line flux profile is not restricted only to the emission from the Weigelt knots. Instead, emission from the entire Weigelt complex is altered, even at distances far away from the measured position of the knots. This effect is shown in more details in Fig.~\ref{fig:profiles2}, for the specific case of [Fe\,{\sc ii}]~$\lambda4815$ and [Fe\,{\sc iii}]~$\lambda4659$ at two different phases.}

\btxt{At phase 0.085, emission from [Fe\,{\sc iii}]~$\lambda4659$ is weak and concentrated around a compact region close to Weigelt C (Fig.~\ref{fig:profiles2}\textit{a}). In contrast, at the same phase, [Fe\,{\sc ii}]~$\lambda4815$ emission is spread all over the Weigelt complex, with two distinctive flux maxima around the Weigelt knots (Fig.~\ref{fig:profiles2}\textit{c}). Later on, at phase 0.516, however, the intensity of emission from [Fe\,{\sc iii}]~$\lambda4659$ is substantially increased, as well as its projected size (Fig.~\ref{fig:profiles2}\textit{b}), whereas the [Fe\,{\sc ii}]~$\lambda4815$ emission region slightly shrinks in size, but the total flux, as shown in Section \ref{flux}, remains constant (this is expected, since more Fe$^{2+}$ is recombining into Fe$^{+}$, which in turn recombines into Fe$^{0}$ keeping the number of Fe$^{+}$ roughly the same).}

\btxt{Since the ionization state of the Weigelt knots is governed by the incident flux from the central source, the changes in the spatial distribution of line emission at phase 0.085 relative to the other phases can be explained by a significant reduction of H-ionizing photons reaching the Weigelt complex at early phases of the spectroscopic cycle. This same conclusion was also obtained by \cite{2005ASPC..332..283H} after analyzing the spectrum of Weigelt D across the 2003.5 spectroscopic event.}

\btxt{In the binary scenario, the decrease of H-ionizing photons at phases close to the spectroscopic event is naturally ascribed to the quenching of the flux of the hot secondary by the primary wind, as the system approaches periastron passage, whereas emission from low ionization potential ions is sustained by the flux from the primary star at all phases \citep{2005ASPC..332..283H,2005A&A...436..945H,2002ApJ...581.1154V}.}

\subsection{Ionization structure\label{sec:astrometry}}
\btxt{In order to determine the position of the Weigelt knots at the different wavelengths, we first used a Gaussian filter with a fixed width of 2 pixels (about the size of the PSF FWHM of 0.11\arcsec) to smooth the iso-velocity images, and then we determined the position of maximum flux of each Weigelt knot for each line listed in Table~\ref{tab:linelist}. With this procedure, we were able to determine both the distance from the central source and the position angle (PA) of each knot. We note, however, that emission from He$^0$ is relatively weak and is blended with the PSF for phases earlier than 0.163. Only at phases later than that it is possible to detect  extended emission associated with the Weigelt knots and reliably measure their position.} Table~\ref{tab:astro} summarizes our results.

\begin{figure}
  \centering
    \includegraphics[width=\linewidth]{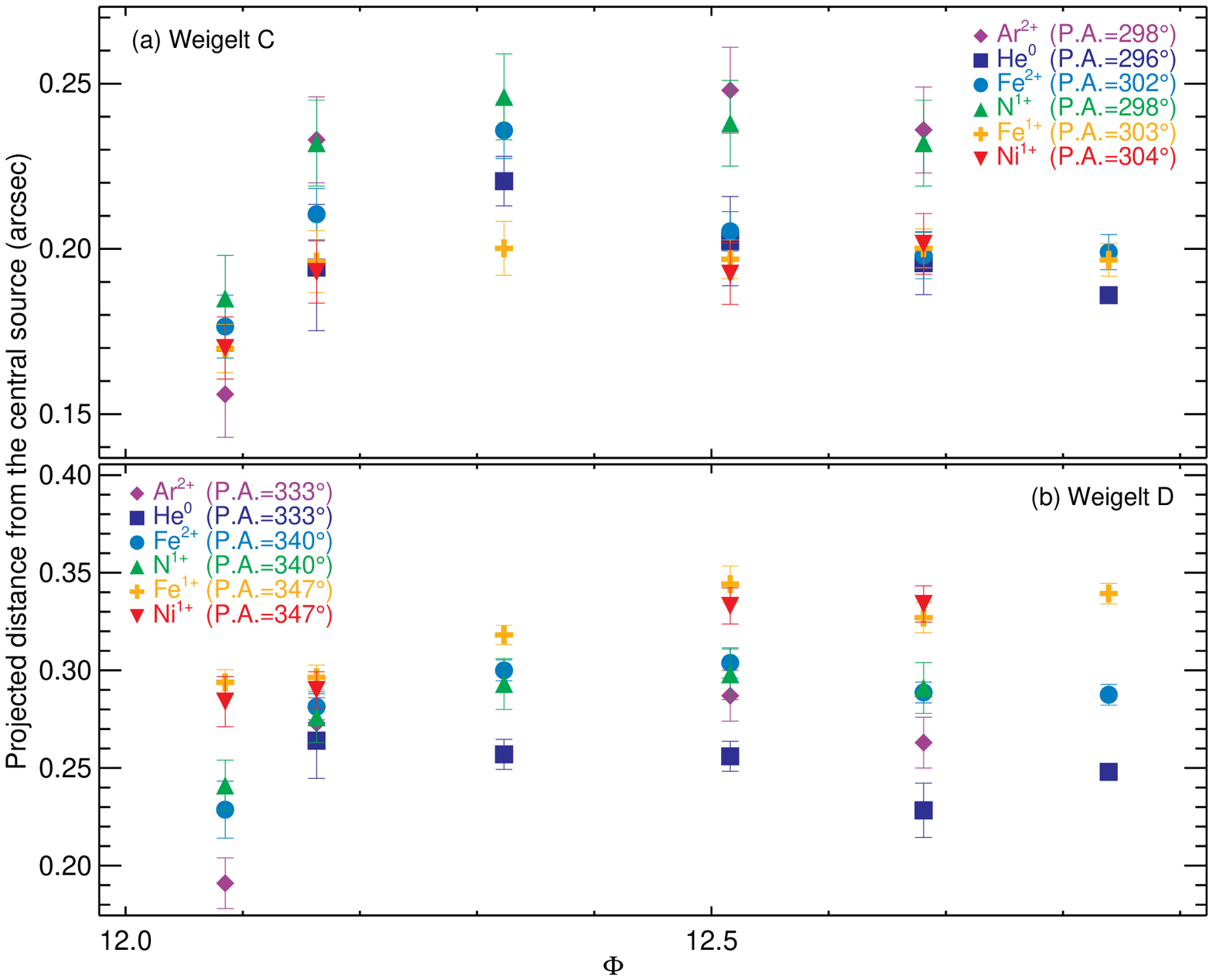}
  \caption{Measured distance of Weigelt C and D from the central source. The observed stratification due to the incident radiation is different for each structures. \btxt{As also reported by \protect \citet{2013ApJ...773...27R}, in Weigelt C}, the high ionization/excitation lines are observed farther away from the central source than the low ionization/excitation ones. The opposite occurs in D. The number in parenthesis indicates the determined PA for the peak of emission for the ion.\label{fig:astrometry}}
\end{figure}

\begin{table*}
    \centering
    \caption{Measured distance from the central source and position angle of the Weigelt knots.\label{tab:astro}}
    \begin{tabular}{lcccccccc}
    
    \hline\hline
    \multirow{2}{*}{Ion} &
    Weigelt &
    \multicolumn{6}{c}{Average distance from the central source (arcsec)} &
    \multirow{2}{*}{PA\textsuperscript{a}}
    \\\cline{3-8}
     &
    knot &
    $\Phi=12.085$ &
    12.163 &
    12.323 &
    12.516 &
    12.680 &
    12.839 & \\\hline\hline
    \multirow{2}{*}{Ar$^{2+}$} & C & 0.156$\pm$0.013 &  0.233$\pm$0.013 & \multirow{2}{*}{no data} &  0.248$\pm$0.013 &  0.236$\pm$0.013 & \multirow{2}{*}{no data} & 298$\degr$ \\
                           & D & 0.191$\pm$0.013 &  0.273$\pm$0.013 &   &  0.287$\pm$0.013 &  0.263$\pm$0.013 &   & 333$\degr$ \\ \cline{2-9}
\multirow{2}{*}{He$^{0}$}  & C & N.\,M.\textsuperscript{b} & 0.194$\pm$0.019 &  0.221$\pm$0.008 &  0.202$\pm$0.013 &  0.196$\pm$0.010 &  0.186$\pm$0.013 & 296$\degr$ \\
                           & D & N.\,M.\textsuperscript{b} & 0.264$\pm$0.019 &  0.257$\pm$0.008 &  0.256$\pm$0.008 &  0.228$\pm$0.014 &  0.248$\pm$0.013 & 333$\degr$ \\ \cline{2-9}
\multirow{2}{*}{Fe$^{2+}$} & C & 0.176$\pm$0.010 &  0.211$\pm$0.008 &  0.236$\pm$0.008 &  0.205$\pm$0.006 &  0.198$\pm$0.007 &  0.199$\pm$0.005 & 302$\degr$ \\
                           & D & 0.229$\pm$0.015 &  0.281$\pm$0.007 &  0.300$\pm$0.005 &  0.304$\pm$0.008 &  0.289$\pm$0.005 &  0.287$\pm$0.005 & 340$\degr$ \\ \cline{2-9}
\multirow{2}{*}{N$^{+}$}  & C & 0.185$\pm$0.013 &  0.232$\pm$0.013 &  0.246$\pm$0.013 &  0.238$\pm$0.013 &  0.232$\pm$0.013 & \multirow{2}{*}{no data} & 298$\degr$ \\
                           & D & 0.241$\pm$0.013 &  0.276$\pm$0.013 &  0.293$\pm$0.013 &  0.298$\pm$0.013 &  0.291$\pm$0.013 &   & 340$\degr$ \\ \cline{2-9}
\multirow{2}{*}{Fe$^{+}$} & C & 0.170$\pm$0.007 &  0.196$\pm$0.009 &  0.200$\pm$0.008 &  0.197$\pm$0.006 &  0.200$\pm$0.006 &  0.197$\pm$0.005 & 303$\degr$ \\
                           & D & 0.294$\pm$0.006 &  0.296$\pm$0.006 &  0.318$\pm$0.005 &  0.344$\pm$0.009 &  0.327$\pm$0.008 &  0.339$\pm$0.005 & 347$\degr$ \\ \cline{2-9}
\multirow{2}{*}{Ni$^{+}$} & C & 0.170$\pm$0.009 &  0.193$\pm$0.009 & \multirow{2}{*}{no data} &  0.193$\pm$0.009 &  0.202$\pm$0.009 & \multirow{2}{*}{no data} & 304$\degr$ \\
                           & D & 0.284$\pm$0.013 &  0.290$\pm$0.009 &   &  0.333$\pm$0.009 &  0.334$\pm$0.009 &  & 347$\degr$ \\\hline\hline
    \multicolumn{7}{l}{\textsuperscript{a}\footnotesize{Average position angle.}} \\
    \multicolumn{7}{l}{\textsuperscript{b}\footnotesize{Not measured due to low line flux and blend with the central source.}}
    
    \end{tabular}
    
\end{table*}

\btxt{The ionization structure of Weigelt C was discussed in detail by \cite{2013ApJ...773...27R}, using HST/STIS data obtained on 2010 March 3 \mbox{($\Phi=12.206$)}. They showed that Weigelt C presents a peculiar ionization stratification, with low ionization potential ions located closer (projected distance) to the central source than high ionization potential ions. Fig.~\ref{fig:astrometry}\textit{a} confirms their results and also shows that both the distance from the central source and the PA at which the maximum flux occurs depends upon the ionization potential of the transition.}

\btxt{The emission from the intermediate ionization potential ions Fe$^{2+}$ and N$^{+}$ is clearly separated, with the former being at a smaller projected distance from the central source than the latter, especially at phases 0.323, 0.516, and 0.680. This is somewhat surprising because the range of ionization energy for these ions is about the same (IP(Fe$^{+}$)=16.2 eV, IP(Fe$^{++}$)=30.7 eV; IP(N$^+$)=14.5 eV, IP(N$^{++}$)=29.6 eV;

 see Table~\ref{tab:linelist}). Therefore, we would expect them to coexist within the same region. Instead, emission from N$^{+}$ is observed at positions closer to that from the ion Ar$^{2+}$ (IP(Ar$^{++}$)=27.6 eV; IP(Ar$^{3+}$)=40.7 eV).}
 
\btxt{Weigelt C does not show any discrepancies for the position of low ionization ions like Fe$^{+}$ (IP+7.9~eV)

and Ni$^{+}$ (IP=7.6~eV);

 they are vary similarly with time. For these ions, the relative increase in the distance between Weigelt C and the central source is about 20\%, whereas for the emission from the high ionization ions, the relative distance is about 3 times larger.}

\btxt{Interestingly, Weigelt D shows a different ionization structure \btxt{than} Weigelt C. In Weigelt D, emission from high ionization potential ions \btxt{are} seen closer to the central source than the low ones (Fig.~\ref{fig:astrometry}\textit{b}). At phases as early as \btxt{0.085}, Weigelt D is already divided into three clearly separate regions of low, intermediate, and high ionization potential ions. The maximum relative increase in the distance from the central source to Weigelt D, measured using the low ionization ions, is about 15\%, whereas for the intermediate and high ionization ions, the relative increase in the distance ranges from 25 to 50\%.}

One remarkable feature not noticed \btxt{in previous works} is that, for both knots, the PA of the maximum flux for a given spectral line depends on the ionization potential. The peak flux position for high ionization ions is observed at smaller PA than low ionization ions. The range in the PA with the ionization potential is smaller for C than for D (see the PA indicated \btxt{in the legends} of Fig.~\ref{fig:astrometry} and Table~\ref{tab:astro}). The dependence of the PA on the ionization potential was not noticed previously as only narrow- ($\approx40$~\AA) or medium-band ($\approx550$~\AA) imaging filters were used to measure the position of the Weigelt knots \citep{1986A&A...163L...5W,1988A&A...203L..21H,1995RMxAC...2...11W,2004AJ....127.1052D,2004ApJ...605..405S,2012ASSL..384..129W}, which does not suffice to differentiate the line from the continuum emission.

\begin{figure}
  \centering
  \includegraphics[width=\linewidth]{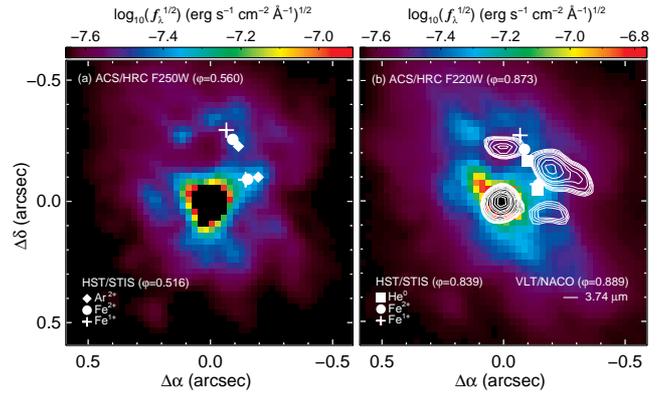}
  \caption{\btxt{ACS/HRC image (F250W and F220W, Proposal ID 10844 \btxt{PI K.~Davidson}) at two different phases: (a) 0.560 and (b) 0.873.} The actual phase ($\varphi$) for each observation is indicated in the labels. Note that the periastron image was obtained on 2002 Oct 14 ($\Phi=10.873$). The symbols mark the position of the peak of emission of the indicated ions. The contours illustrate the distribution of bright infrared structures detected on NACO imaging obtained on 2002 Nov 16 \citep[$\Phi=10.889$;][]{2005A&A...435.1043C} using the NB\_374 filter (central wavelength around 3.74~$\mu$m; dominated by emission from the hydrogen line Pf$\gamma$). The ACS/HRC and the NACO images were obtained just one month apart. Regions of low near-UV flux correspond to the location of the bright infrared structures. The peak of line emission is observed around these bright infrared structures, similar to the near-UV flux, suggesting that the optical Weigelt knots are, actually, UV-illuminated (and consequently ionized) parts of the infrared bright structures. North is up and East is to the left.\label{fig:naco}}
\end{figure}

The narrow-band imaging filters F631N and F658N, often used to map the Weigelt knots, include contributions from continuum and line emission from [S\,{\sc iii}]~$\lambda6313$ and [N\,{\sc ii}]~$\lambda6585$\footnote{\cite{2004ApJ...605..405S} estimate that at most $15$\% of the total flux is due to line emission.}, respectively, causing the position of the optical knots to be biased towards the position of maximum emission from intermediate to high ionization ions. Moreover, the spectroscopic studies performed with single slit sampled only specific position angles with the slit centered upon $\eta$~Car. In this work, we can differentiate both spatially and spectroscopically and focus only on the emission from the Weigelt complex, excluding high velocity structures.

In Fig.~\ref{fig:naco}, we compare our maps to the distribution of bright infrared structures around the central source as detected by \cite{2005A&A...435.1043C}. For Weigelt D, the position of the peak in the line emission for the ions used in this work is at the edges of the bright northern structure dubbed D$^\prime$. Low ionization potential ions are always located at positions with higher PA than the high or intermediate ionization potential peaks. Also, the infrared structure (Weigelt D$^\prime$) is spatially coincident with the low near-UV flux region, suggesting that the line emission and the near-UV flux are both delineating the edges of a large structure seen only at wavelengths longer than the optical. This suggests that the Weigelt knots seen in the optical are not a physical structure but an illumination effect caused by the incident UV radiation from the central source illuminating the outer parts of the infrared bright structures.

\subsection{Proper motion and formation date}
\btxt{The changes in the ionization structure of the Weigelt knots led us to wonder if they have any influence on the measurements of proper motion as done in previous work \citep[\textit{e.g.}][]{1995RMxAC...2...11W,2004AJ....127.1052D,2004ApJ...605..405S,2012ASSL..384..129W}. This is important because changes in the ionization structure can lead to variations in the measured position of the knots as seen in HST images, resulting in misleading ejection dates. This is especially true if observations are not obtained at the same phase within each orbital cycle.}

\begin{figure}
  \centering
  \includegraphics[width=\linewidth]{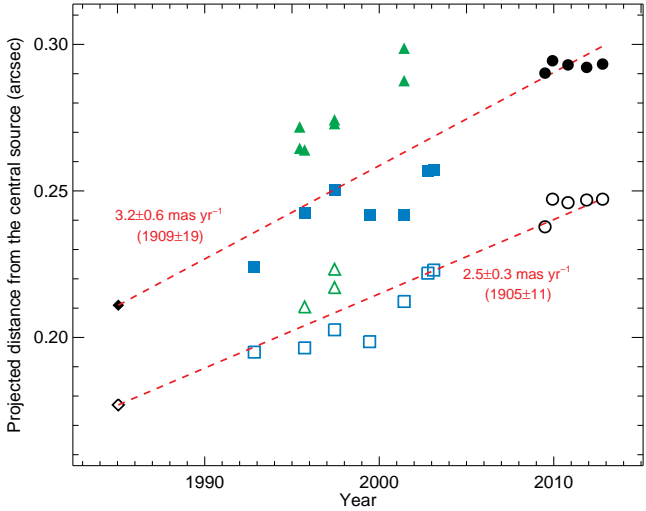}
  \caption{\btxt{Proper motion and ejection date of Weigelt C (open symbols) and D (filled symbols).  Projected distance from the central source as measured by \protect\cite{1988A&A...203L..21H} (diamonds), \protect \cite{2004AJ....127.1052D} (triangles), \protect \cite{2004ApJ...605..405S} (squares), and this work (circles). \label {fig:proper_motion} }}\end{figure}

\begin{table*}
    \centering
    \caption{De-reddened brightness of the Weigelt knots. The observed flux was corrected using $A_V=2$ and $R_V=4$.\label{tab:flux}}
    \begin{tabular}{rccccccc}
    
    \hline\hline

    \multirow{2}{*}{Spectral line} &
    Weigelt &
    \multicolumn{6}{c}{$I_{0\lambda}$ ($10^{-10}$ erg\,s$^{-1}$\,cm$^{-2}$\,\AA$^{-1}$\,arcsec$^{-2}$)}
    \\\cline{3-8}
    &
    knot ID &
    $\Phi=12.085$ &
    12.163 &
    12.323 &
    12.516 &
    12.680 &
    12.839 \\\hline
\multirow{2}{*}{[Ar III]~$\lambda7137$} & C & 0.322$\pm0.004$ & 0.965$\pm0.005$  & \multirow{2}{*}{no data}  & 1.628$\pm0.010$  & 1.956$\pm0.012$  & \multirow{2}{*}{no data} \\
                                        & D & 0.101$\pm0.001$ & 0.656$\pm0.002$  &     & 1.256$\pm0.008$  & 1.344$\pm0.008$  &  \\ \cline{2-8}
\multirow{2}{*}{He I~$\lambda4714$}     & C & 0.145$\pm0.003$ & 0.309$\pm0.004$  & 0.360$\pm0.004$  & 0.429$\pm0.005$  & 0.442$\pm0.005$  & 0.461$\pm0.006$    \\
                                        & D & 0.050$\pm0.001$ & 0.120$\pm0.001$  & 0.170$\pm0.001$  & 0.202$\pm0.001$  & 0.208$\pm0.001$  & 0.183$\pm0.001$    \\ \cline{2-8}
\multirow{2}{*}{[Fe III]~$\lambda4659$} & C & 0.644$\pm0.007$ & 1.205$\pm0.007$  & 1.300$\pm0.004$  & 1.458$\pm0.008$  & 1.748$\pm0.012$  & 1.893$\pm0.013$    \\
                                        & D & 0.316$\pm0.001$ & 0.833$\pm0.002$  & 1.085$\pm0.004$  & 1.287$\pm0.007$  & 1.407$\pm0.009$  & 1.319$\pm0.008$    \\ \cline{2-8}
\multirow{2}{*}{[N II]~$\lambda5756$}   & C & 3.868$\pm0.039$ & 6.373$\pm0.031$ & 6.625$\pm0.023$ & 7.067$\pm0.030$ & 8.202$\pm0.045$ & \multirow{2}{*}{no data}  \\
                                        & D & 1.988$\pm0.010$ & 4.423$\pm0.015$  & 5.250$\pm0.029$  & 5.893$\pm0.028$  & 6.202$\pm0.035$  &  \\\cline{2-8}
\multirow{2}{*}{[Fe II]~$\lambda4815$}  & C & 1.900$\pm0.017$ & 1.880$\pm0.013$  & 1.842$\pm0.013$  & 1.767$\pm0.014$  & 1.748$\pm0.012$  & 1.962$\pm0.015$    \\
                                        & D & 1.943$\pm0.017$ & 1.748$\pm0.010$  & 1.741$\pm0.010$  & 1.836$\pm0.011$  & 1.868$\pm0.014$  & 2.044$\pm0.018$    \\ \cline{2-8}
\multirow{2}{*}{[Ni II]~$\lambda7413$}  & C & 1.022$\pm0.007$ & 1.028$\pm0.007$  & \multirow{2}{*}{no data} & 1.028$\pm0.008$  & 1.047$\pm0.007$  & \multirow{2}{*}{no data}  \\
                                        & D & 0.946$\pm0.007$ & 0.946$\pm0.005$  &     & 0.984$\pm0.006$  & 1.010$\pm0.006$  &  \\\hline\hline

    \end{tabular}

\end{table*}

\begin{figure}
  \centering
    \includegraphics[width=\linewidth]{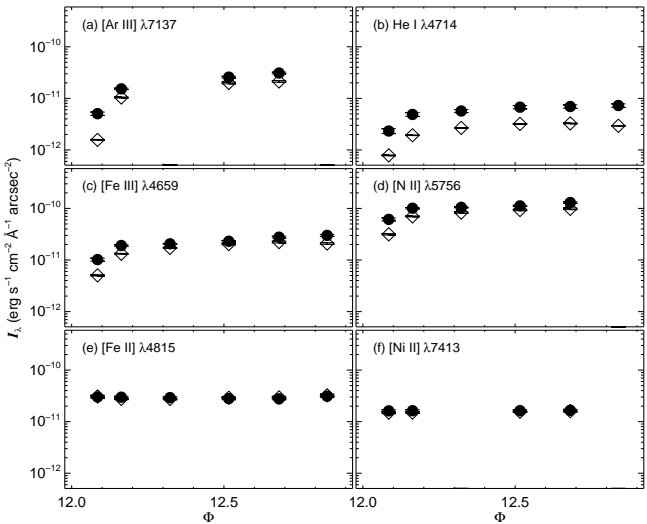}
  \caption{Absolute (a--f) and relative (g--l) brightness variations in the Weigelt complex. The filled circles correspond to Weigelt C and the open diamonds, to Weigelt D. Note that the absolute brightness was not corrected by extinction. Instead, the de-reddened brightness for each line is listed in Table~\ref{tab:flux}.\label{fig:flux}}
\end{figure}

In order to evaluate the influence of the ionization changes in the proper motion measurements, we used data from \cite{1988A&A...203L..21H}, \cite{2004AJ....127.1052D}, and \cite{2004ApJ...605..405S} to compare with the measurements presented in this work. The results are shown in Fig.~\ref{fig:proper_motion}.

\btxt{First, we determined the fractional variation of the [N\,{\sc ii}]~$\lambda5756$ line emission in our dataset for each phase and then subtracted the corresponding amount from the position of the knots as measured using HST narrow images. Without any correction due to ionization effects, we derived a proper motion of \mbox{$\dot{\theta}_\rmn{C}=2.3\pm0.6$} and \mbox{$\dot{\theta}_\rmn{D}=3.1\pm0.8$~mas\,yr$^{-1}$} for Weigelt C and D, respectively. After accounting for changes in the ionization structure, we derived proper motions of \mbox{$\dot{\theta}_\rmn{C}=2.5\pm0.3$} and \mbox{$\dot{\theta}_\rmn{D}=3.2\pm0.6$~mas\,yr$^{-1}$} for Weigelt C and D, respectively. This is still consistent with the previous results, indicating that changes in the ionization structure have small influence over the position of the structures measured using HST narrow filter. Note, however, that for spectral images obtained at phases $\varphi<0.15$ or $\varphi>0.85$, it is important to correct for the changes in the ionization structure, since at these phases the position of the knots as measured by intermediate or high ionization potential ions can change by significant amounts.}

\btxt{The derived average epoch of formation of the Weigelt structures is \mbox{$1907$ ($\pm16$~yr)}, which is in accordance with $1908$ ($\pm12$~yr) and $1880$ ($\pm20$~yr) obtained by \cite{2004ApJ...605..405S} and \cite{2012ASSL..384..129W}, respectively. Also, considering only measurements from \cite{1988A&A...203L..21H} and \cite{2004ApJ...605..405S} (less discrepancy between measurements), results in \mbox{$\dot{\theta}_\rmn{C}=2.2\pm0.3$}, \mbox{$\dot{\theta}_\rmn{D}=2.5\pm0.4$~mas\,yr$^{-1}$}, and an epoch of formation of \mbox{$1902$ ($\pm15$)}. It should be noticed that radiative erosion would mimic a faster proper motion, in which case dates earlier than 1907 are plausible.}

Although our results do not agree with the average value of $1934.1$ derived by \cite{2004AJ....127.1052D}, we note that their  lower and upper limits  ($31.7$ and $16$~yr, respectively) encompasse our values.

\subsection{Line ratio diagnostics}

\btxt{We derived the brightness, electron density and temperature, and dilution factor for the structures forming the Weigelt complex using the atomic models described in \citet{Bautista:2004fa} for Ni$^{+}$, \citet{2000ApJ...544..581B} for He$^{0}$, \citet{1995A&A...293..953Z} and \citet{Bautista:1996tc} for Fe$^{+}$, and \citet{2010ApJ...718L.189B} for Fe$^{2+}$.}\subsubsection{Brightness}\label{flux}

\btxt{The measured brightness was de-reddened by an average total extinction \mbox{$A_V=2.0$; R$_V=4.$} \cite[][]{1995AJ....109.1784D,1999ASPC..179..116H} and the results are presented in Table~\ref{tab:flux}.}

Weigelt C is brighter than D during most of the orbital cycle for emission from intermediate or high ionization potential ions (Fig.~\ref{fig:flux}\textit{a--d}). This is especially true for phases \btxt{0.085, 0.163} and for the entire light curve from He$^{0}$ transitions (this particular case will be addressed in details in section~\ref{sec:temp}). On the other hand, line emission coming from low ionization potential ions do not show significant changes in brightness throughout the cycle (Fig~\ref{fig:flux}\textit{e--f}).

Fig.~\ref{fig:flux}\textit{g--l} shows the line brightness variations across the orbital cycle normalized to the line brightness observed at $\Phi=12.516$. When we look at the emission from intermediate and high ionization potential ions, Weigelt D shows the largest variation relative to apastron. \btxt{Between phases 0.085 and 0.516}, the brightness of Weigelt D increases faster than that of C. \btxt{After phase 0.516}, the brightness of Weigelt D begins to fade whereas that of C continues to increase (at least until $\varphi=0.839$).

\begin{figure*}
  \centering

  	\subfloat[]{\includegraphics[width=0.49\linewidth]{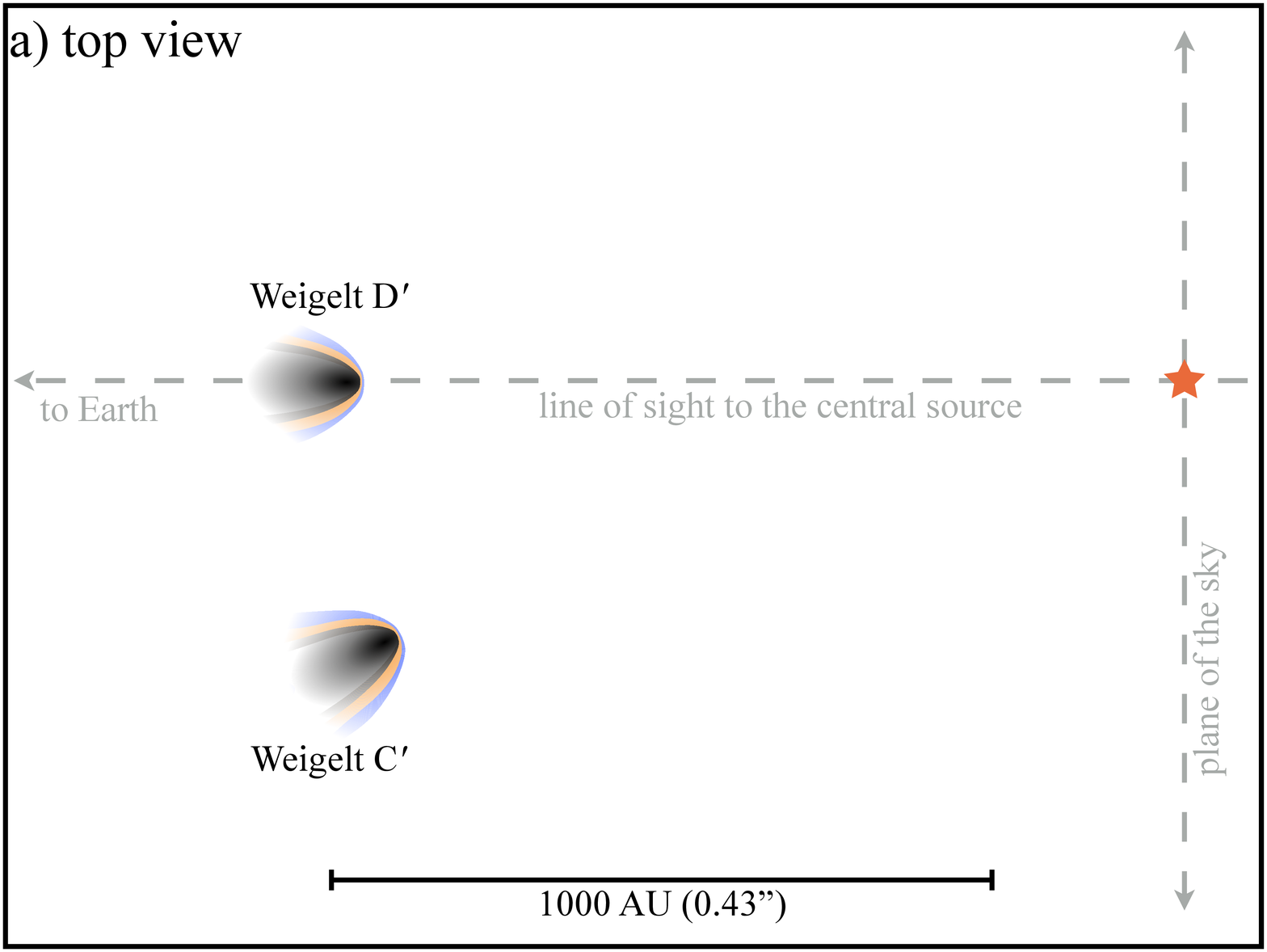}}
  	\subfloat[]{\includegraphics[width=0.49\linewidth]{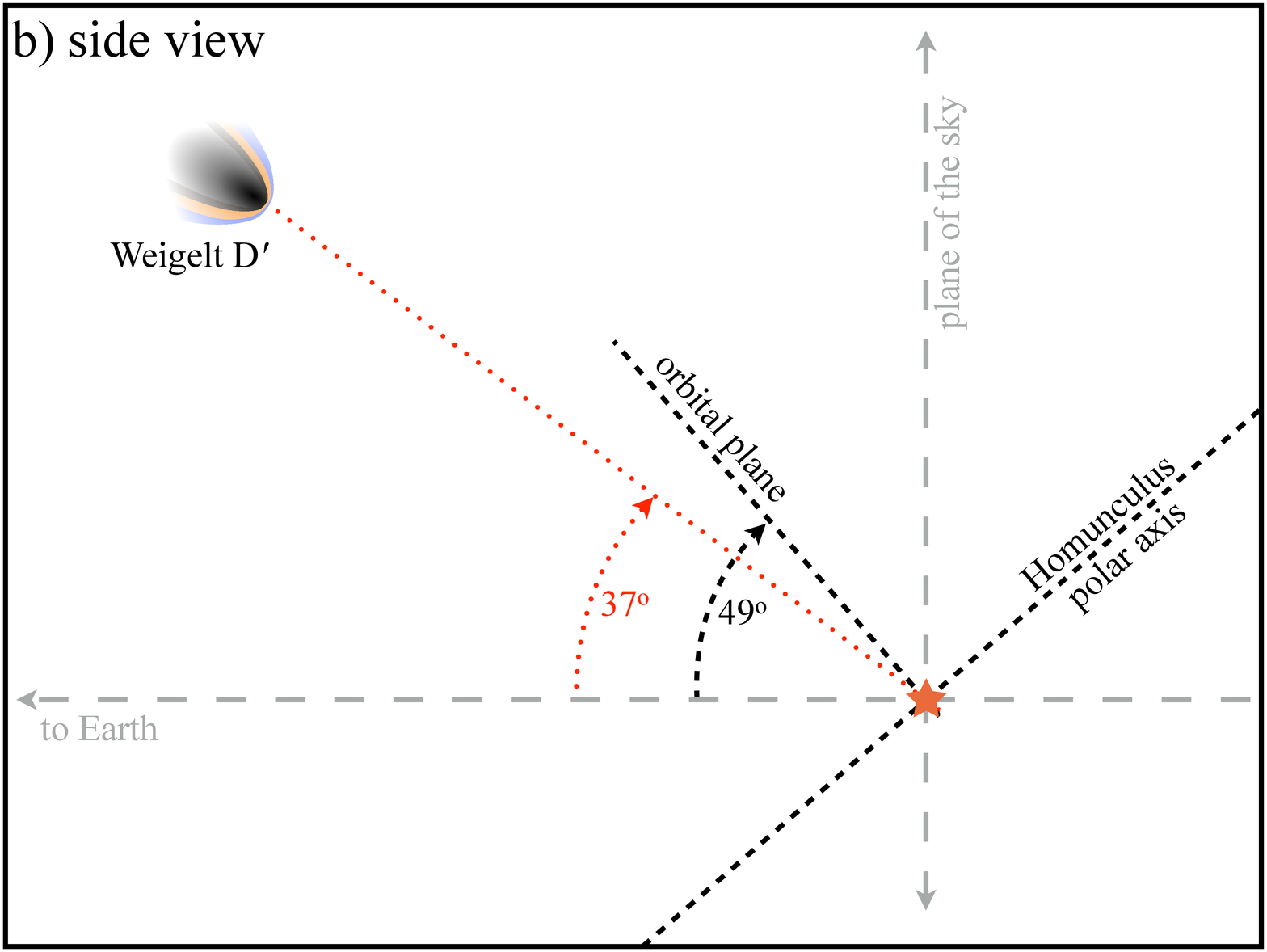}}

  \caption{\btxt{Relative orientation of the Weigelt knots. (a) View perpendicular to the plane containing the Weigelt knots and the central source. (b) View perpendicular to the plane formed by Weigelt D$^\prime$, the central source, and the line of sight of the observer. The shape of the knots and the photo-evaporating regions is a depiction and shall not be taken as strictly correct, since the actual shape of the structures are unknown.}\label{fig:sketch}}
\end{figure*}

The low ionization potential lines do not show significant changes in the brightness for either knot, which is expected since even the primary star (cooler than the secondary) can maintain the population of the low energy levels of these ions. However, we note that emission from Fe$^{+}$ does show a small increase (about 10\%) in the brightness of both knots \btxt{at phases \mbox{$\varphi$=0.085, 0.163, and 0.839}}. \btxt{At these phases}, the size of the low ionization emitting region increases as the higher ionized plasma recombines (cf. Fig.~\ref{fig:profiles2}\textit{b}).

\btxt{At early phases ($\varphi=0.085$ and 0.163), Weigelt C becomes brighter than D by a factor of \mbox{2--3} in the light of the forbidden transitions from high/intermediate ionization potential ions (see Table~\ref{tab:flux}). This variation can be explained if most of the illuminated face of the near-infrared bright structure, Weigelt D$^\prime$, is turned away from us due to its orientation regarding our line of sight and the central source. This scenario is illustrated in Fig.~\ref{fig:sketch} where the inclination of $37\degr$ relative to the line of sight was obtained based on the distance between the structures and the central source derived from the determination of the dilution factor (see section~\ref{sec:df}). In this configuration, Weigelt D$^\prime$ would be almost aligned with the line of sight to the central source, and the optical emission lines that we observe would be formed both on the ablated borders of the near-infrared structure (where mostly of the lines from the intermediate and high ionization potential ions are formed) and the diffuse radiation field that forms around and behind it (where lines from the low ionization potential ions are formed).}

In other words, it seems that we are looking at Weigelt D$^\prime$ from behind it (or in conjunction regarding the primary star), whereas the emission lines would be formed in the ionized halo around the borders of this structure invisible in the optical, but detectable in the near-infrared region. Thus, any intrinsic changes in the illumination of the knots by the central source would be converted to a higher rate of variation in the brightness of Weigelt D$^\prime$ than for Weigelt C$^\prime$ simply because the projected area (as seen by the observer) where the lines are formed, is smaller in the former than in the latter.

Therefore, it is likely that the differences between the brightness of the knots is a result of their relative orientation regarding the observer's line of sight to the central source, and not to an intrinsic reduced ionizing flux impinging upon Weigelt D$^\prime$ only.

\subsubsection{Electron density}\label{sec:dens}

We compared the observations with the results from atomic models for Fe$^{+}$ and Fe$^{2+}$ to infer the local physical conditions in the Weigelt knots. 

\begin{figure}
  \centering
  \includegraphics[width=\linewidth]{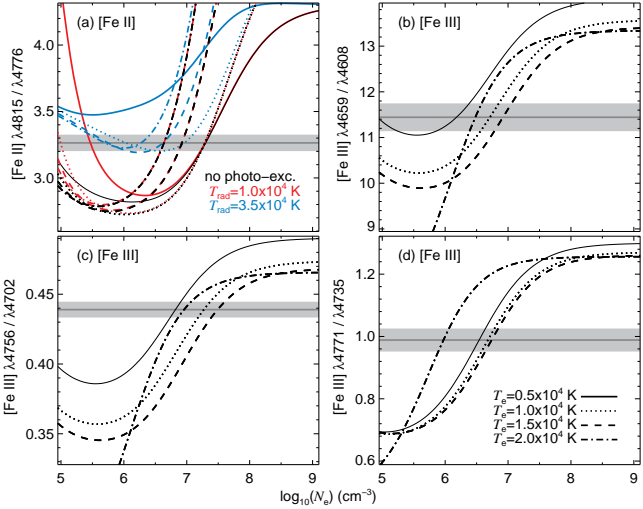}
  \caption{Time-averaged observed line ratios from (a) Fe$^{+}$ and (b--d) Fe$^{2+}$. The horizontal solid line is the observed line ratio averaged over the interval $\Phi=12.085-12.839$, whereas the shaded area corresponds to its absolute error. For the [Fe\,{\sc ii}]~$\lambda4815/4776$ line ratio in (a), we also show the effect of photo-excitation for a dilution factor $\log{w}=-9$ and two different radiation temperatures $T_\rmn{rad}=10^4$~K and $3.5\times10^4$~K. Each curve corresponds to a different electron temperature as indicated in (d).\label{fig:fe_ratios}}
\end{figure}

Our Fe$^{+}$ model accounts for photo-excitation due to the external radiation field from the central source. The total fluxes for a selected group of density-sensitive transitions from Fe$^{+}$ and Fe$^{2+}$ were measured within the same regions mentioned in the previous section. We calculated the ratio between the total fluxes for each pair of transitions listed in Table~\ref{tab:lineratios} and determined the corresponding range of allowed electron densities. The observed line ratios (averaged over the orbital cycle) and the results from our atomic models for a set of electron temperatures are shown in Fig.~\ref{fig:fe_ratios}. The derived range of allowed electron densities for each line ratio is shown in Table~\ref{tab:lineratios}.

Due to their low ionization potential energy (only 7.9~eV above ground state), in addition to collisional excitation/de-excitation, Fe$^{+}$ ions can also have their low-lying level population modified by photo-excitation due to an external radiation field. This is obviously the case for the Weigelt complex where $\eta$~Car A and B provide a variable and relatively strong UV radiation field.

During most of the orbital cycle, the Weigelt complex is illuminated by both stars. The primary provides a radiation field with radiation temperature $T_\rmn{rad}\approx10^4$~K whereas the secondary is responsible for the high temperature radiation, $T_\rmn{rad}\approx3.5\times10^4$~K. Close to periastron passages, however, the secondary gets enshrouded by the dense primary wind, causing the radiation temperature to drop to $T_\rmn{rad}\approx10^4$~K (the temperature in the outer parts of the primary wind). Thus, in the Weigelt complex, the population of the low-lying levels of Fe$^{+}$ ions is expected to vary according to the characteristic temperature of the incident radiation. The effect of photo-excitation on the [Fe\,{\sc ii}]~$\lambda4815/\lambda4776$ line ratio, when the radiation temperature changes from $T_\rmn{rad}=10^4$~K to $3.5\times10^4$~K, is illustrated in Fig.~\ref{fig:fe_ratios}\textit{a}.

On one hand, at $T_\rmn{rad}=10^4$~K, photo-excitation processes have little influence on the line ratio for electron densities $\log{N_\rmn{e}}>6$, as compared to calculations without photo-excitation. Even at electron temperatures as low as $T_\rmn{e}=5000$~K, in the high electron density regime, the level population calculated including photo-excitation processes does not differ from that without it. In other words, since the critical density for [Fe\,{\sc ii}]~$\lambda4815$ and [Fe\,{\sc ii}]~$\lambda4776$ is $\sim10^7$~cm$^{-3}$, under a relatively weak radiation field and local densities $\log{N_\rmn{e}}>6$ the line emission is mainly due to collisional processes.

On the other hand, at $T_\rmn{rad}=3.5\times10^4$~K, the low-lying levels might be substantially affected by photo-excitation. However, the importance of this processes will be ultimately defined by the local electron density and temperature. For high temperatures ($T_\rmn{e}>10^4$~K) and densities ($\log{N_\rmn{e}}>7$), collisional processes dominate over radiative transitions and the level population will be set by collisions. For low electron temperatures ($T_\rmn{e}<10^4$~K) and local densities $<10^9$~cm$^{-3}$, radiative processes are an important source in populating the levels.

\begin{figure}
  \centering
  \includegraphics[width=\linewidth]{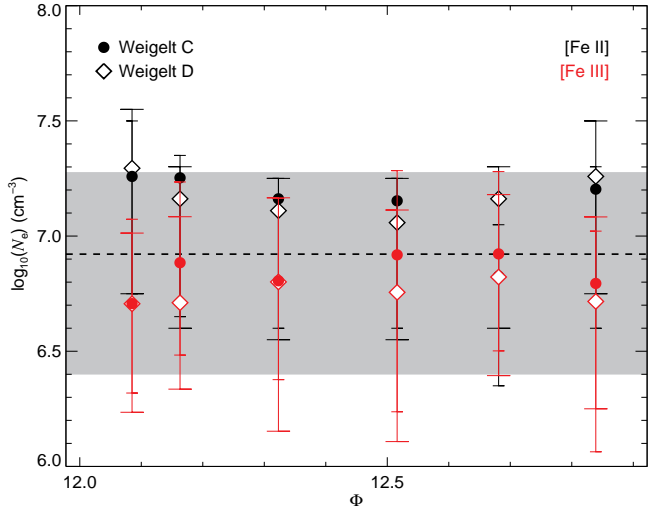}
  \caption{Electron density over the orbital cycle. The symbols indicate the average electron density for Weigelt C (circles) and D (diamonds) derived using line ratios from Fe$^{+}$ (red) and Fe$^{2+}$ (black) ions. The horizontal dashed line indicates the time-averaged electron density using both Fe$^{+}$ and Fe$^{2+}$ (as obtained from Fig.~\ref{fig:fe_ratios}). The gray area shows the range of allowed values for $n_\rmn{e}$ based on the time-averaged measurements for both ions shown in Fig.~\ref{fig:fe_ratios}. The slightly asymmetry in the uncertainties towards low electron densities is due to the slope of the line ratio curve, which is shallower in that part than for high electron densities.\label{fig:dens_phase}}
\end{figure}

\begin{table}
    \centering
    \caption{Density sensitive line ratios.\label{tab:lineratios}}
    \begin{tabular}{ccccc}

    \hline\hline
    
    \multirow{2}{*}{Ion} &
    \multirow{2}{*}{Line ratio} &
    Observed &
    $\log{N_e}$\textsuperscript{b} &
    $\langle \log{N_e} \rangle$\textsuperscript{c}
    \\
    &
    &
    ratio\textsuperscript{a} &
    (cm$^{-3}$) &
    (cm$^{-3}$) \\\hline
Fe$^{+}$ & $\lambda$4815/$\lambda$4775 & 3.265\,$\pm$\,0.062  & $6.6-7.4$ & $7.2^{+0.2}_{-0.6}$ \\ 
\cline{2-4}
\multirow{3}{*}{Fe$^{2+}$} & $\lambda$4659/$\lambda$4608 & 11.451\,$\pm$\,0.304 & $5.9-7.1$ & \multirow{3}{*}{$6.8^{+0.4}_{-0.7}$} \\
			& $\lambda$4756/$\lambda$4702 & 0.439\,$\pm$\,0.006  & $6.8-7.6$ & \\ 
			& $\lambda$4771/$\lambda$4735 & 0.989\,$\pm$\,0.037  & $5.9-6.9$ & \\\hline
    \multicolumn{4}{r}{Average density\textsuperscript{d}} & $6.9^{+0.4}_{-0.6}$ \\\hline\hline
    \multicolumn{5}{l}{\textsuperscript{a}\footnotesize{Time-averaged line ratio.}} \\
    \multicolumn{5}{l}{\textsuperscript{b}\footnotesize{Range of allowed time-averaged densities.}} \\
    \multicolumn{5}{l}{\textsuperscript{c}\footnotesize{Average density for each ion.}} \\
    \multicolumn{5}{l}{\textsuperscript{d}\footnotesize{Based on the average density for each ion.}} \\

    \end{tabular}
    \end{table}

\begin{figure}
  \centering
  \includegraphics[width=\linewidth]{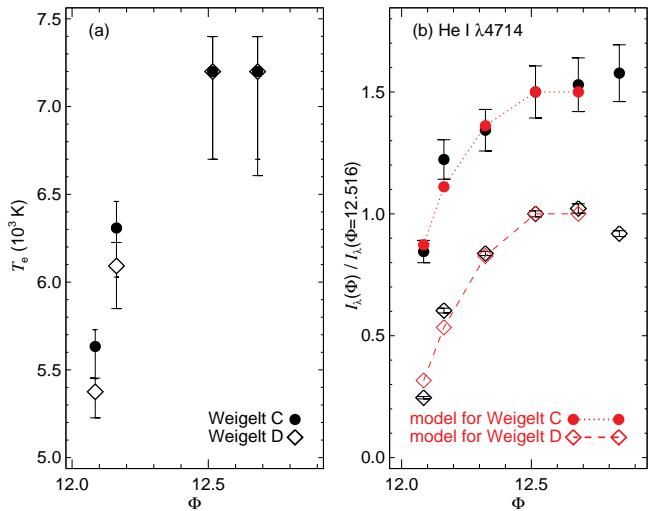}
  \caption{(a) Electron temperature over the orbital cycle, for each knot, as derived from the  He\,{\sc i}~$\lambda7283/\lambda7065$ line ratio. (b) Comparison between the observed (black symbols) and predicted (connected red symbols) variations in the specific intensity of  He\,{\sc i}~$\lambda4714$. We used a fixed electron density of $10^{6.9}$~cm$^{-3}$ and the temperatures shown in (a) to calculate the brightness of the line at each epoch relative to apastron ($\Phi=12.516$). For $\Phi=12.323$, we used an interpolated value for the electron temperature. Since we do not have data for  He\,{\sc i}~$\lambda7065$ or  He\,{\sc i}~$\lambda7283$ at $\Phi=12.839$, we did not attempt to model the brightness at this epoch. Measurements for Weigelt C were shifted by 0.5 in the ordinate axis.\label{fig:temp_phase}}
\end{figure}

Therefore, we expect the line ratio [Fe\,{\sc ii}]~$\lambda4815/\lambda4776$ to vary across the orbital cycle according to the characteristic temperature of the incident radiation. Indeed, the derived electron density does show a small, yet clearly detectable, variation as a function of the phase of the spectroscopic cycle (Fig.~\ref{fig:dens_phase}). During most of the orbital cycle ($12.1<\Phi<12.8$), the electron density, as derived from the [Fe\,{\sc ii}]~$\lambda4815/\lambda4776$ line ratio, is essentially constant and lower than that determined at phases \btxt{0.085 and 0.163}. This is a direct result of changing the temperature of the incident radiation. The small amplitude (less than 0.15 dex) of the variation in the electron density across the orbital cycle implies that radiative processes are comparable to collisional processes, which for the observed line ratio would only occur for electron temperatures $T_\rmn{e}>10^4$~K.

For the electron density diagnosis using Fe$^{2+}$ transitions, we used three suitable line ratios (Fig.~\ref{fig:fe_ratios}\textit{b}--\textit{d}). At each phase, the electron density was derived by considering the average value after cross-correlating the allowed range of densities from each line ratio. The result is also shown in Fig.~\ref{fig:dens_phase}. As opposite to Fe$^{+}$, Fe$^{2+}$ ions are not sensitive to the radiation field and, therefore, do not show any trend or systematic variation across the orbital cycle.

\subsubsection{Electron temperature}\label{sec:temp}
In order to probe the electron temperature of the Weigelt complex, we used the ratio between He$^0$ lines arising from singlet and triplet states. This method relies on the difference between the electron impact excitation coefficient rate for each He$^0$ state, which varies as a function of the electron temperature \citep{2001PhPl....8.5303B}. For electron densities $n_e\lesssim 10^{11}$~cm$^{-3}$, this method results in relatively accurate diagnostics for the plasma temperature.

The He\,{\sc i}~$\lambda7283$/$\lambda7065$ line ratio was chosen for many reasons. Since the lines are in the red part of the optical window and are close to each other, they are relatively less susceptible to extinction effects. Furthermore, both lines have the same branching ratio and are strong in the spectrum of the Weigelt knots, since they arise from $n=3$. The only potential problem with this line ratio is the close proximity of the $2\,^3S$ meta-stable level. Since the optical depth for a line transition is proportional to the population of the lower level, the large population of the meta-stable level increases the optical depth for the $2\,^3S-\,n\,^3P$ transitions \citep{2002ApJ...569..288B}, leading to important radiative transfer effects.

\begin{figure}
  \centering
  \includegraphics[width=\linewidth]{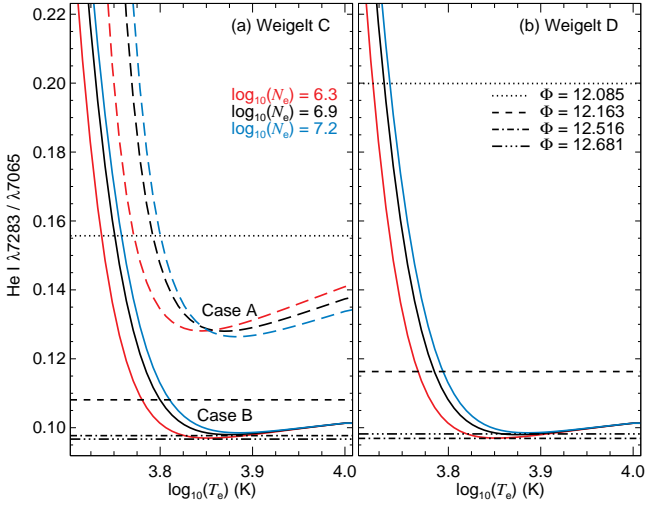}
  \caption{Electron temperature diagnostic from the line ratio He\,{\sc i}~$\lambda7283/\lambda7065$. Three different electron densities ($\log{N_\rmn{e}}=6.3$, 6.9, and 7.2) were used to calculate the line ratio curves shown here (dashed and solid lines correspond to optically thin and thick regimes, respectively). The horizontal lines in both panels correspond to the observed ratio for each phase of the spectroscopic cycle as labelled in the right panel.\label{fig:temp_ratio}}
\end{figure}

For the triplet line of interest in our analysis -- He\,{\sc i}~$\lambda7065$ ($2\,^3P-\,3\,^3S$) -- there are two optical depth effects that may play a role in an optically thick plasma. 
First, a large optical depth for the He\,{\sc i}~$\lambda3889$ ($2\,^3S-3\,^3P$) can increase the $\lambda7065$ line intensity by converting $\lambda3889$ photons into $\lambda4.3$~$\mu$m plus $\lambda7065$ plus $\lambda10830$ photons. 
Second, a large optical depth for $2\,^1P$ - 1$^1$S would increase the population of the  2$^1$P  level by orders of magnitude; thus collisional transitions would populate the 2$^3$P level, leading to re-absorption of $\lambda7065$ photons and thus a decrease in the observed line intensity. 
The net result of these optical depth effects is, however, difficult to estimate and, in order to be adequately studied, would require full three-dimensional radiative transfer modeling, which is out of the scope of the present work. Here, we just assume, for simplicity, that the net result does not considerably affect the He\,{\sc i}~$\lambda7065$ line intensity.

We set up He\,{\sc i} atomic models for a plasma in two different regimes: optically thin (Case A) and optically thick (Case B) to radiation originating from transitions to $n=1$. The atomic models were set for three electron densities (minimum, average, and maximum value, as determined in the previous section; see Fig.~\ref{fig:temp_ratio}) for each Weigelt knot. Naturally, the average electron temperature was determined using the curve for the average electron density, whereas the uncertainty in $T_\rmn{e}$ was determined by the curves for the minimum and maximum $n_\rmn{e}$ (Fig.~\ref{fig:temp_phase}\textit{a}).

The optically thin regime is not a good approximation for either knot at $\phi>0.1$, for which the observations can only be reproduced by an optically thick plasma (Fig.~\ref{fig:temp_ratio}). This is  fact expected; given that the Weigelt complex has a rich spectrum of low ionization species (\textit{e.g.} Fe$^{+}$), most of the helium content must be neutral (He$^+$/He$^0\approx0$), resulting in large optical depths.

\begin{figure}
  \centering
  \includegraphics[width=\linewidth]{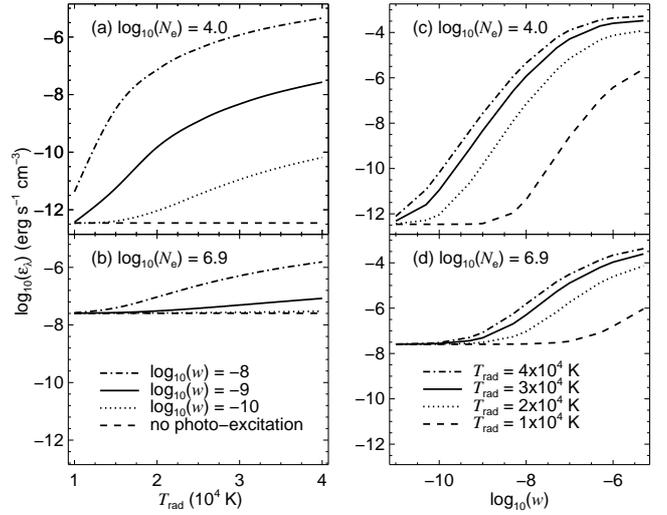}
  \caption{[Fe~{\sc ii}]~$\lambda4641$ line emissivity as a function of the radiation temperature (left panel) and dilution factor (right panel). The upper panels show the variations in the line emissivity in a relatively low density plasma ($10^4$~cm$^{-3}$), whereas the lower panels are for a high density one ($10^{6.9}$~cm$^{-3}$). Note that, for example, changing $w$ from zero (no photo-excitation) to $10^{-8}$ causes the line emissivity from a high density plasma to be enhanced by a factor of $10^2$, for a star with $T_\rmn{eff}=4\times10^4$~K. In the low density plasma, the same change in $w$ would result in a change in the emissivity by a factor of $10^7$. For these calculations, we assumed an electron temperature of $T_\rmn{e}=6000$~K.\label{fig:photo}}
\end{figure}

The changes in the electron temperature over the orbital cycle are shown in Fig.~\ref{fig:temp_phase}\textit{a}. The absolute error in the measured ratio is about 1 per cent. The observed ratios for phases \btxt{0.516 and 0.681} are close to the local minimum of the line ratio models for Case B, which results in large uncertainties for the determined temperatures. Nevertheless, it is apparent that Weigelt C has a higher electron temperature than D \btxt{at phases 0.085 and 0.163}. The difference is not large but is sufficient to be clearly detected at \btxt{these} phases. This also can be explained by the orientation of the knots relative to our line of sight to the central source. If Weigelt D is in inferior conjunction, then most of its ionized region is blocked from our view, and only part (closer to the central source) of the relatively cool borders of the structures will contribute to the observed line flux. Weigelt C would be apparently hotter than D in the case where we can see a relatively large fraction of the ionized region (the morphology of Weigelt C could also play an important role here).

In order to check the reliability of the determined electron temperatures, we attempted to reproduce the observed normalized brightness of the He\,{\sc i}~$\lambda4714$ line by providing the atomic models with the derived $T_\rmn{e}$ for each phase of the spectroscopic cycle. Thus, we solved the equations for level population at each electron temperature and used the following relation\footnote{Demonstrated in Appendix~\ref{sec:appendix}} to determine the normalized intensity of the line
\begin{equation}
 \frac{I_{\lambda}(\Phi)}{I_{\lambda}(\Phi_0)} = \frac{\left(\frac{N_i}{N_j}\right)_{\Phi}}{\left(\frac{N_i}{N_j}\right)_{\Phi_0}},
 \label{eq:normalintensity}
\end{equation}
where $\Phi_0$ (=12.516) is the epoch chosen as the normalization factor, and thus, the left-hand side of the equation is the normalized brightness of the line, whereas the right-hand side is the relative change (due to $T_\rmn{e}$) in the population of the levels involved in the transition. The results are shown in Fig.~\ref{fig:temp_phase}\textit{b}, where it can be seen that the observations can be reproduced by the derived electron temperatures at a fixed electron density of $10^{6.9}$~cm$^{-3}$ (the average derived density).

\subsubsection{Distance from the central source\label{sec:df}}
Since both ions, Fe$^{+}$ and Ni$^{+}$, have almost the same low ionization potential (less than 8~eV), the strong UV radiation field associated with $\eta$~Car A and B can alter the distribution of the energy levels population via photo-excitation. This is especially true in low density nebulae, where photo-excitations can significantly contribute to the line emissivity. Even at high density, where collisional processes dominate, photo-excitations can still affect the line emissivity, depending on the luminosity of the radiation source.

\begin{figure}
  \centering
  \includegraphics[width=\linewidth]{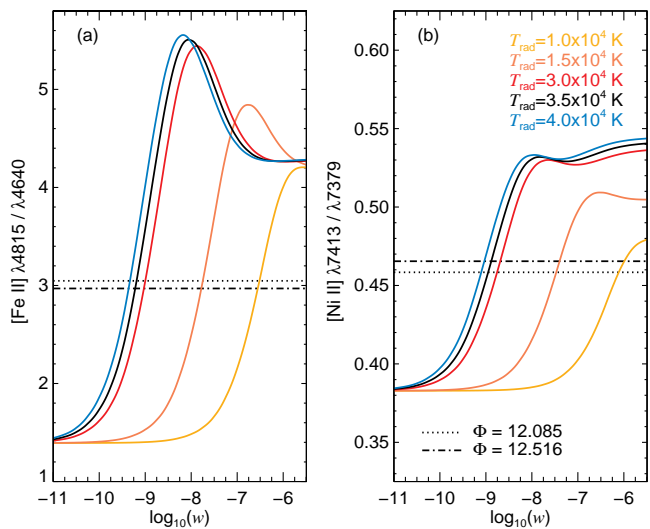}
  \caption{Dilution factor plots for $n_\rmn{e}=10^{6.9}$~cm$^{-3}$, $T_\rmn{e}=6000$~K, and five different radiation temperatures, $T_\rmn{rad}$. For an exciting/ionizing source with $T_\rmn{rad}=3.5\times10^4$~K (secondary star), we have an average $\log{w}=-9.1$, or $r/R_*=8472$. If the exciting/ionizing source has $T_\rmn{rad}=10^4$~K ($T_\rmn{rad}=1.5\times10^4$~K), then the average $\log{w}=-6.3$ ($\log{w}=-7.6$) would correspond to $r/R_*=353$ ($r/R_*=1578$).\label{fig:dfactor2}}
\end{figure}

It is crucial, therefore, to take into account the contributions from the background radiation photo-excitations to the level population. This is usually achieved by assuming that the background radiation source radiates like a blackbody, in which case the mean intensity of the radiation field is given by
\begin{equation}
  J_\lambda = \frac{1}{2}\left(1-\sqrt{1-\frac{R_*^2}{r^2}}\right)\,B_\lambda(T_{\rmn{rad}}),
\end{equation}
where $J_\lambda$ is the mean intensity at a distance $r$ from a radiation source that has a radius $R_*$, and $B_\lambda(T_{\rmn{rad}})$ is the spectral radiance of a blackbody at temperature $T_{\rmn{rad}}$. Therefore, by parametrizing the mean intensity as $J_\lambda=w\,B_\lambda(T_{\rmn{rad}})$, we introduce the dilution factor $w$, defined as (for large distances)
\begin{equation}
  w = \frac{1}{4} \left(\frac{R_*}{r}\right)^2.
  \label{eq:w}
\end{equation}
Thus, by simply providing a temperature and a dilution factor, the photo-excitation contributions can be evaluated and accounted for within atomic models. As an example, the effect of including photo-excitation on the calculation of the [Fe\,{\sc ii}]~$\lambda4641$ line emissivity\footnote{Defined as $\varepsilon_\lambda = \Delta E\, A_{ij}\, n_i$, where $\Delta E$ is the energy difference between the upper level $i$ and the lower level $j$, $A_{ij}$ is the spontaneous decay coefficient, and $n_i$ is the number density of the upper level.} as well as its dependence with $T_{\rmn{eff}}$ and $w$ are illustrated in Fig.~\ref{fig:photo}. Thus, by determining the dilution factor, one can infer the distance to the radiation source.

For the Weigelt complex, we used two specific line ratios to probe the dilution factor. Namely, [Fe\,{\sc ii}]~$\lambda4815/\lambda4640$ and [Ni\,{\sc ii}]~$\lambda7413/\lambda7379$. The former was chosen because [Fe\,{\sc ii}]~$\lambda4815$ responds differently than [Fe\,{\sc ii}]~$\lambda4640$ for variations in the dilution factor. Both lines are sensitive to photo-excitation, but [Fe\,{\sc ii}]~$\lambda4640$ is less sensitive than \mbox{[Fe\,{\sc ii}]~$\lambda4815$} for dilution factors $\log{w}<-8.5$. As an example, a plasma with $\log{N_\rmn{e}}=6.9$ and $T_\rmn{e}=6000$~K being illuminated by a star with $T_\rmn{eff}=4\times10^4$~K, will have the line emissivity of [Fe\,{\sc ii}]~$\lambda4640$ increased by a factor of 10 while that of [Fe\,{\sc ii}]~$\lambda4815$ will increase by a factor of 200, when the dilution factor changes from $10^{-11}$ to $10^{-8.5}$.

As for the [Ni\,{\sc ii}]~$\lambda7413/\lambda7379$, this is a well-known line ratio \citep[][and references therein]{1995A&A...294..555L,1996ApJ...460..372B,2006MNRAS.370.1991B} for which the $\lambda7379$ line emission is enhanced by continuum UV pumping. The first step of this process occurs via the strong dipole transition from the ground state $a\,^2D_{5/2}$ to the opposite odd parity level $z\,^2D^0_{5/2}$ ($\lambda1742$). It is followed by spontaneous decay to $a\,^2F_{7/2}$ via another UV transition at $2279$~\AA, which thereby enhances the population of the upper level that gives rise to the [Ni\,{\sc ii}]~$\lambda7379$ line.

Since the radiation from either star can be responsible for the photo-excitation of the low lying levels of Fe$^{+}$ and Ni$^{+}$, we employed atomic models with different $T_\rmn{rad}$ for a fixed  electron density and temperature of $\log{N_\rmn{e}}=-6.9$ and $T_\rmn{e}=6000$~K, respectively. Note that these are the time-averaged values. Based on previous work, the effective temperature of the secondary star should be in the range $3-4\times10^4$~K \citep{2005ApJ...624..973V,2008MNRAS.387..564T,2010ApJ...710..729M}. As for the primary, the effective temperature (where $\tau_\rmn{Ross}=2/3$) should be around $10^4$~K due to its extended, optically thick LBV wind \citep{2001ApJ...553..837H,2012MNRAS.423.1623G}. The results are shown in Fig.~\ref{fig:dfactor2}, where we present the observed line ratio \btxt{at phase 0.085 and 0.516}. It is evident, as expected, that they do not change significantly.

On one hand, the observed line ratios suggest an average dilution factor of about $\log{w}=-9.1$ for $T_\rmn{rad}=3.5\times10^4$~K, which would correspond to the secondary star. On the other hand, if the background radiation source is ascribed to the primary star, with $T_\rmn{rad}=10^4$~K, then $\log{w}=-6.3$. From Eq.~\ref{eq:w}, a $\log{w}=-9.1$ corresponds to $r/R_*=8472$, whereas a $\log{w}=-6.3$ implies $r/R_*=353$. Thus, the distance of the Weigelt complex to the central source can be estimated if we adopt a reasonable value for the radius of the star responsible for the photo-excitation. 

X-ray observations and 3D hydrodynamics simulations suggest that the secondary star has a radius of about $30$~R$_\odot$ \citep{2007RMxAC..30...29C,2008MNRAS.388L..39O,2009MNRAS.394.1758P,2012MNRAS.420.2064M}. If we assume that this is a fair classification for the secondary star and that it is responsible for the background radiation during most of the orbital period, then it would imply that the Weigelt complex is about \mbox{$1020$~AU} from the central source. Alternatively, if the primary is the main source of the UV continuum radiation, then assuming a radius of \mbox{$R(\tau_\rmn{Ross}=2/3)=890$~R$_\odot$} \citep{2001ApJ...553..837H} would correspond to a distance of \mbox{$1460$~AU} to the Weigelt complex. Either way, given the uncertainties in the stellar radii, we can say that the results are in good agreement, and an average value of \mbox{$1240$ ($\pm220$)~AU} can be assumed. A caveat here is that the primary star might not be well represented by a blackbody due to strong absorption by Fe$^{+}$ transitions in the stellar wind.

\begin{figure}
  \centering
  \includegraphics[width=\linewidth]{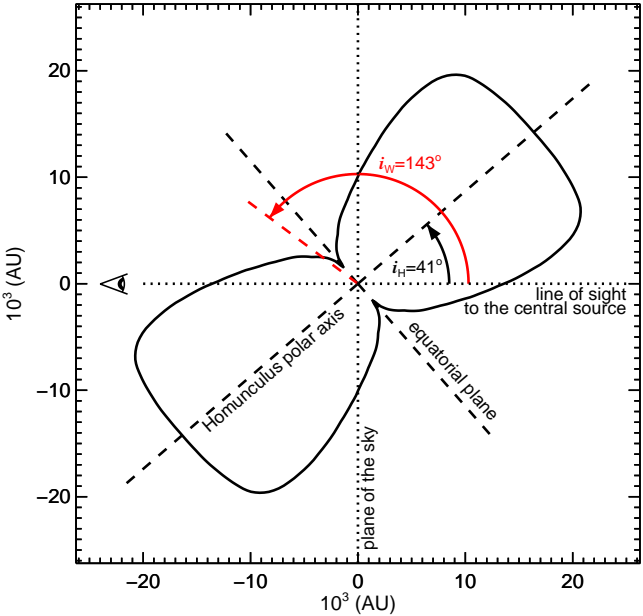}
  \caption{Orientation of the Weigelt structures. The Homunculus nebula has an inclination angle of $i_\rmn{H}=41\degr$, whereas the Weigelt complex has $i_\rmn{W}=143\degr$. We adopted the geometry and distance to the Homunculus nebula derived by \protect \cite{2006ApJ...644.1151S}. The observer is looking at the system from the left.\label{fig:geometry}}
\end{figure}

At a distance of 2350~pc \cite[][]{2006ApJ...644.1151S}, the average projected position of Weigelt D of $0.316$ ($\pm0.026$)~arcsec (as measured by Fe$^{+}$ and Ni$^{+}$ emission lines) would correspond to a projected distance of \mbox{$743$ ($\pm61$)~AU}. Thus, the angle between Weigelt D and the line of sight to the observer, as seen from the central source, is about $37\degr$ ($\pm12\degr$), which implies $i_\rmn{W}=143\degr$ ($\pm12\degr$) (see Fig.~\ref{fig:geometry} for the definition of the inclination angle, $i_\rmn{W}$). This result confirms that the Weigelt complex is indeed located in the equatorial region (and near the binary orbital plane, as well). Again, the large error in the calculations are due to uncertainties in the stellar radii.

\section{Discussion}\label{sec:discussion}

\subsection{The optical Weigelt knot B}
Since several previous observations always reported the observations of three optical structures \citep[\textit{e.g.}][]{1986A&A...163L...5W,1988A&A...203L..21H,1995RMxAC...2...11W,2004ApJ...605..405S,2012ASSL..384..129W} and we only detected two of them (C and D) in the \textit{HST/STIS} data, one might wonder where knot B is. From our analysis of the specific intensity throughout the Weigelt complex, we only detected two regions that significantly contribute to the total flux at velocities around $-40$~km\,s$^{-1}$, namely, Weigelt C and D. We did not detect any structure that could possibly be associated to the optical knot B. What could have happened to it?

One possibility is that this structure has been at least partially photo-evaporated by the strong radiation pressure and stellar wind from the central source. However, given the estimated mass of these knots, $\sim10^{-3}-10^{-2}$~M$_\odot$ \citep{2003AJ....125.1458S,1994ApJ...422..626H,1995AJ....109.1784D}, this process would require a sustained unrealistic high mass-loss rate in the range \mbox{$10^{-5}-10^{-4}$~M$_\odot$\,yr$^{-1}$}, over the period $1988-2012$, in order to completely photo-evaporate a structure like this. The situation gets even worse considering that narrow band H alpha images (filter width = 4 nm) taken in 2008 show a knot brighter than Weigelt D that can be associated to knot B \citep[see figure 1\textit{d} of][]{2012ASSL..384..129W}. This suggests that photo-evaporation is not a realistic explanation.

 A rapid decrease of the illumination of Weigelt D$^\prime$ at the position of the optical knot B, and/or obscuration by Weigelt D$^\prime$ in our line of sight, offers a simpler explanation for the observed brightness drop. This effect is also observed in young stellar objects (YSOs). For example, the knots in Hubble's variable nebula (surrounding the Herbig Be star R~Mon) can change their brightness within months \citep{1989MNRAS.239..665L,1989MNRAS.237..621S}. In the case of the Weigelt complex, it is evident that the physical structure cannot change so fast because of the large size and distances involved, but   illumination or obscuration can change as fast as observed. 

Given the much more extended Weigelt D$^\prime$ structure, a combination of photo-evaporation, illumination and obscuration might be the more plausible explanation of the apparent disappearance of Weigelt B. An independent observation at radio wavelengths, where dust extinction is nil, of emission of a hydrogen recombination line might demonstrate that Weigelt B still exists, but is obscured to us at visible wavelengths.
\subsection{The connection between the Weigelt structures and the Butterfly Nebula}
Apart from the distance to the central source, the age and kinematics of the Weigelt complex are within the range of values observed for the `Butterfly Nebula' (hereafter BN), a region composed by slow-moving clumps of material distributed around the equatorial region \citep{2005A&A...435.1043C,2011AJ....141..202A}. 

The age distribution of the BN components indicates that approximately 40\% of the clumps were formed between 1840 and 1870 (during the Great Eruption), while about 30\% were formed between 1880 and 1910 (coincident with the Lesser Eruption of 1890). Most of the clumps are moving radially outward at space velocities between 50 and 150~km\,s$^{-1}$ (with a mode of 85 km\,s$^{-1}$) and are located between 1500 and 4000~AU from the central source \citep{2011AJ....141..202A}.

Regarding the distance to the central source, there is only one structure of the BN that seems to be located at a distance similiar to the Weigelt complex: the \mbox{SE clump}. This structure lies on the opposite side of the central source, suggesting that it could be the SE counterpart of the Weigelt complex. Also, an intriguing connection arises when we consider the orientation of the binary orbit \citep{2007ApJ...660..669N,2008MNRAS.388L..39O,2009MNRAS.394.1758P,2012MNRAS.420.2064M}; the Weigelt complex and the SE clump seems to be aligned with the direction of the semi-major axis. Could the binary nature of the central source be responsible for the observed geometry of the BN?

For a binary system, the gravitational force (far away from any libration point) will have a maximum intensity along the axis joining both stars. Thus, material closer to the orbital plane will be subject to a stronger gravitational pull than material at high stellar latitudes. Hence, for low eccentricity orbits, the time-averaged effect of the gravitational force over a distribution of material around the binary system will be isotropic. In other words, the change in the momentum of the circumstellar material will be isotropic and, therefore, it will move away from the central source with the same velocity, forming a ring around the binary system.

A long-period, high eccentricity orbit, as in the $\eta$~Car system, implies that during most of the time the stars will be aligned along the semi-major axis, which would result in a greater time-averaged change in the momentum along this direction than at other directions. The net result would be a departure from the ring geometry, where now the material along the semi-major axis is moving slower and is located closer to the binary system than the material at any other direction. Note that this result does not depend on the longitude of periastron, but only on the relative position of the stars.

For a system like $\eta$~Car, assuming that the primary has $90$~M$_\odot$, the secondary has $30$~M$_\odot$, and they are about 30~AU apart during apastron, one can show that the gravitational force will be about two times more intense along the semi-major axis than perpendicular to it, implying a time-averaged change in momentum by the same factor. Thus, if there were no substantial changes in the mass of the structures forming the BN since its formation, one should expect that the distance of the Weigelt complex and the SE clump to the central source would be about two times smaller than for material at positions like the SW or NE region \citep[see \mbox{figure 7} of][]{2005A&A...435.1043C}. The space velocity should also differ by the same amount.

That is exactly what the observations indicate; the average distance of the Weigelt complex and the SE clump to the central source ($\approx 1700$~AU) is about half of that measured for the SW and NE region ($\approx 3300$~AU). About the same factor is also observed for the space velocity; the Weigelt complex and SE clump are moving at an average speed of $\approx 60$~km\,s$^{-1}$, whereas the SW and NE regions are moving at $\approx 100$~km\,s$^{-1}$ \citep[note, however, that this value does not adequately sample the entire SW region since Doppler velocity measurements were only made for the SW clump; see figure 10 of][]{2011AJ....141..202A}. Therefore, the geometry of the BN might be an indication that the binary nature of the central source could have had a strong influence on shaping it.

\subsection{The equatorial material as remnants of the Lesser Eruption}
The ages and velocities we derive in this work for the Weigelt structures suggest that they are correlated to the BN and were formed during the outburst that formed the Little Homunculus \citep{2003AJ....125.3222I,2005MNRAS.357.1330S}. The age distribution of the BN suggest that 70\% of it was formed during the epoch of the two outbursts; the first one that formed the Homunculus and the second one associated with the Little Homunculus. But how can such energetic events produce slow-moving clumps as we observe in the BN?

Evidently, if the star was an RSG or YHG when the equatorial material was ejected, then relatively low velocities are anticipated due to the low escape velocity of stars in these evolutionary phases. But the ages derived for the equatorial material seems to indicate that they were ejected during the outbursts and, thus, should be moving with much higher velocities (like the two bipolar nebulae). It is possible that, right after being ejected by an eruptive event, the major contribution to changes in the momentum of the clumps would be due to gravitational forces. Hence, most of the energy acquired from the stellar outburst would have been spent as work to drive the clump away from the central source.

There are other two forces that will also act upon the clump. One is the radiative force due to the radiation field of the stars and the other is the drag force due to the stellar wind related to the mass-loss rate. Both forces will push the clump away from the central source, in opposition to the gravitational pull. Assuming that the clump expands as it moves outward so as to keep a constant solid angle, the gravitational force will rapidly decrease to the point it will be superseded by the drag and radiative forces, which will then accelerate the clump. The equation of motion for such clump is
\begin{equation}
    \frac{dv}{dt} = -a_\rmn{g} + a_\rmn{w} + a_\rmn{r}.
    \label{eq:motion}
\end{equation}
The first term of this equation is the gravitational acceleration, given by
\begin{equation}
    a_\rmn{g} = \frac{G\,M_*}{r^2},
\end{equation}
where $M_*$ is the mass of the star and $r$ is the distance from the clump to the central source. The second term of Eq.~\ref{eq:motion} corresponds to the drag caused by the stellar wind, which is given by
\begin{equation}
    a_\rmn{w} = \frac{C_D\,\rho_\rmn{w}(r)\,(v_\rmn{w}(r)-v_\rmn{c})^2\,\pi R_\rmn{c}^2}{2M_\rmn{c}},
\end{equation}
where $C_D$ is the drag coefficient, $\rho_\rmn{w}(r)$ and $v_\rmn{w}(r)$ are, respectively, the wind density (obtained from the mass-loss rate) and velocity (a simple $\beta$-law) at a distance $r$ of the central source, $v_\rmn{c}$ and $R_\rmn{c}$ are the clump velocity and radius (both depend on $r$), respectively, and  $M_\rmn{c}$ is the clump mass. Finally, the third term of Eq.~\ref{eq:motion} is the radiative acceleration due to the central source, which, for simplicity\footnote{A more realistic approach would have to adopt the CAK theory for the radiative acceleration due to both lines and electron scattering continuum, which will be addressed in a forthcoming paper (Teodoro et al.).}, is given by
\begin{equation}
    a_\rmn{r} = \frac{R_\rmn{c}^2}{M_\rmn{c}} \frac{L_*}{4\,c\,r^2},
\end{equation}
where $L_*$ is the luminosity of the central source. In this equation, the clump is considered optically thick. Therefore, by estimating the contribution of each term of Eq.~\ref{eq:motion}, we can derive the full trajectory of a typical clump formed near the central source.

\begin{figure}
  \centering
  \includegraphics[width=\linewidth]{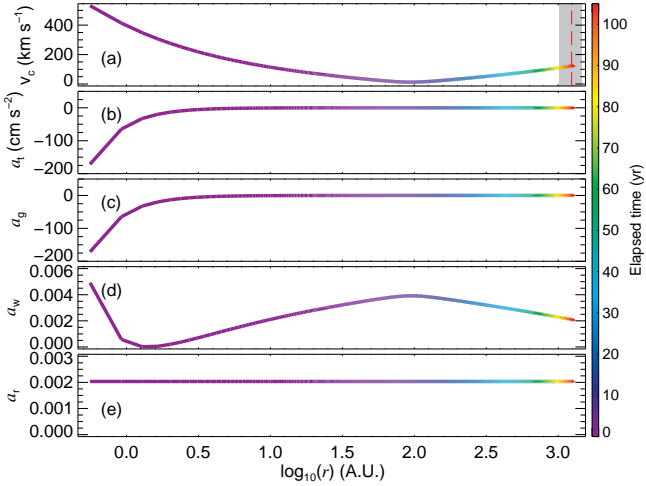}
  \caption{(a) Velocity and (b) total acceleration as a function of distance from the central source for a clump like Weigelt D. The vertical dashed line in (a) is the average distance of the Weigelt knots to the central source as derived in this paper (\mbox{1240 AU}; the shaded area is the uncertainty). The total acceleration is the net result of the combination of (c) gravitational pull, (d)  drag force due to the stellar wind, and (e) radiative acceleration. For this simulation, we assumed a clump mass of $1.2\times10^{-3}$~M$_\odot$. The colors indicate the time since ejection and goes from zero to \mbox{105 yr} (the derived age of the Weigelt complex). \label{fig:clump}}
\end{figure}

As a proof of concept -- that slow-moving material can be formed in eruptions like the one that formed the Little Homunculus --, consider a single star like the primary \citep[$90~\rmn{M}_\odot$, $10^{6.7}~\rmn{L}_\odot$, and $R_*=60~\rmn{R}_\odot$;][]{2001ApJ...553..837H} that has been through an outburst similar to the one that formed the Little Homunculus. The total kinetic energy of the Little Homunculus \citep[$v=250$~km\,s$^{-1}$ and $m=0.1$~M$_\odot$;][]{2005MNRAS.357.1330S} is about $6.2\times10^{46}$~erg. If we assume that about $7$\% of this energy was intercepted by a clump like \mbox{Weigelt D$^\prime$}, then the initial velocity of the clump at a distance of about 0.56~AU\footnote{We are assuming that such clumps are formed in the stellar wind. How they are formed, though, is beyond the scope of the present work.} (corresponding to $2\rmn{R}_*$) would be about 530~km\,s$^{-1}$ (see Fig.~\ref{fig:clump}\textit{a}).

Owing to the strong gravitational potential, the clump will be decelerated to approximately 2\% of the initial value in less than 20 yr after the ejection. Meanwhile, the clump travelled about 100~AU. At this point, the gravitational force acting on the clump will be less than the combined contributions from radiation and stellar wind, and thus the clump begins to accelerate.

About 105~yr after the ejection, the clump will be at 1240~AU from the central source, moving at a speed of approximately $125$~km\,s$^{-1}$ (Fig.~\ref{fig:clump}\textit{a}), which is approximately the parameters estimated for Weigelt D$^\prime$. We did not consider the binary nature of the central source in this calculation, which would have influenced the final velocity of the clump by contributing mainly with gravitational and drag forces. Thus, slow-moving material can, in principle, be formed even in energetic events like the ones experienced by the central source of $\eta$~Car.

At first sight, an accelerated motion for the equatorial material may be contradictory, since most of the studies done to establish the age of the nebular features by analysis of their proper motion have found a uniform motion. Nevertheless, this apparent contradiction can be explained by noting that substantial acceleration of the equatorial material should have occurred in a relatively short period of time, just after the ejection of the material. For the model discussed in this section, Fig.~\ref{fig:clump}\textit{b} shows that the total acceleration of the clump drops from $-200$~cm\,s$^{-2}$ to almost zero about \mbox{16 yr} after its ejection, primarily due to the rapidly decrease in the gravitational pull (Fig.~\ref{fig:clump}\textit{c}). Then, the drag and radiative acceleration becomes dominant for the rest of the time, but the combined components have a magnitude of only $10^{-3}$~cm\,s$^{-2}$ (Fig.~\ref{fig:clump}\textit{d-e}). Hence, the assumption of constant motion is valid for most of the clump's lifetime.

Note, however, that we are not affirming that the mechanism describe above is the only explanation for the equatorial structure. Our point is that it is possible to form slow-moving structures even in high-energy events, like those experienced by $\eta$~Car, and without any specific requirement for the eruptive star (\textit{e.g.} that it would be in the RSG phase). Further detailed studies, especially at infrared and radio wavelengths, are still required to settle the true origin of the equatorial material.

\section{Conclusions}\label{sec:conclusions}
We have presented the first full mapping of the Weigelt complex using \textit{HST/STIS} data obtained at different phases of the spectroscopic cycle. The results of our analysis are summarized as follows.

\begin{itemize}
\item The optical Weigelt knots C and D are extended structures with different sizes, shapes, and orientation regarding our line of sight to the central source. The actual extent of these structures is only revealed by mapping the spatial distribution of line emission from different ions, as we have done here;

\item As suggested by previous work in the infrared region and corroborated by the present work, the \textit{optical} Weigelt knots are associated with the ionized region formed around the actual structures, where material is being photo-evaporated;

\item The position of the optical Weigelt knots, as measured using imaging data, is the result of the contribution of each spectral feature to the flux inside the filter used to observe them; since most of the optical images taken with \textit{HST} are dominated by continuum and/or line emission from intermediate or high ionization potential ions (\textit{e.g.} [S\,{\sc iii}]~$\lambda6313$ is the brightest line inside filter F631N), the spatial location of the optical Weigelt knots is coincident with our measured position of the peak of emission from ions like Ar$^{2+}$, He$^0$, Fe$^{2+}$, and N$^{+}$;

\item Weigelt knots B and D are two different illuminated parts of the same structure: the infrared bright structure called Weigelt D$^\prime$;

\item For the ionized region around Weigelt D$^\prime$, the peak of line emission from intermediate and high ionization potential ions are located closer to the central source and at smaller PAs than line emission from low ionization potential ions. Regarding the position of the peak of line emission, the opposite is observed around Weigelt C$^\prime$;

\item Both structures show a dependence of the PA on the ionization energy in the sense that the peak line emission from low ionization potential ions is located at larger PAs than intermediate or high ones. This is more conspicuous around Weigelt D$^\prime$ than around C$^\prime$;

\item The difference in the ionization structure of the ionized region around Weigelt C$^\prime$ and D$^\prime$ is due to their position, shape, and orientation relative to our line of sight;

\item Weigelt C is brighter than D regarding line emission from intermediate or high ionization potential ions;

\item From Fe$^{2+}$ line ratio analysis, we have derived an average total extinction of $A_V=2.0\pm0.5$ (adopting $R_V=4.0$);

\item From Fe$^{+}$ and Fe$^{2+}$ emission line ratios, we have derived an electron density of $\log{N_\rmn{e}}=6.9\pm0.5$;

\item We have shown that the electron temperature in the Weigelt structures, as inferred from the  He\,{\sc i}~$\lambda7283/\lambda7065$, varies from $T_\rmn{e}=5500$ to $7200$~K \btxt{as the system goes from phase 0.085 to 0.516}, respectively;

\item Depending on the source of the background radiation responsible for photo-excitation, the dilution factor for the Weigelt structures is $\log{w}=-9.1$ for $T_\rmn{rad}$ in the range \mbox{$3$--$4\times10^4$~K} and $\log{w}=-6.3$ for $T_\rmn{rad}=10^4$~K;

\item Confirming previous estimates, the Weigelt complex indeed lies close to the equatorial plane (about $12\degr$ away from it). The average distance from the central source to the Weigelt structures is about \mbox{1240~AU}. These results suggest that the Weigelt structures are related to the `Butterfly Nebula' in nature.
\end{itemize}

\section*{Acknowledgments}
M.\,T. is supported by CNPq/MCT-Brazil through grant 201978/2012-1. This research has made extensive use of NASA's Astrophysics Data System, \href{http://idlastro.gsfc.nasa.gov}{{\sc idl} Astronomy User's Library}, and David Fanning's \href{http://www.idlcoyote.com/index.html}{{\sc idl} Coyote library}. We dedicate this work to Dr. O.~Chesneau for his important contributions to many areas of Astronomy.

\bibliographystyle{mnras}
\bibliography{refs_20141107.bib}

\appendix

\section{Specific intensity of a spectral line}\label{sec:appendix}
From the formal solution of the transfer equation, we have
\begin{equation}
  I_{\nu}(\tau_{\nu}) = I_{\nu}(0) \, e^{-\tau_{\nu}} + S_{\nu} \, (1 - e^{-\tau_{\nu}}),
\end{equation}
where $\tau_{\nu}$ is the monochromatic optical depth, $I_{\nu}(\tau_{\nu})$ is the specific intensity (or brightness) of the line at $\tau_{\nu}$ (therefore, $I_{\nu}(0)$ is the brightness of the incident radiation at $\tau_{\nu}=0$), and $S_{\nu}$ is the source function of the medium. Assuming that $I_{\nu}(0)\approx0$, we have
\begin{equation}
  I_{\nu}(\tau_{\nu}) \approx S_{\nu} \, (1 - e^{-\tau_{\nu}}).
  \label{eq:transfer1}
\end{equation}
Recalling that, in the optically thick regime ($\tau_{\nu}\gg1$), $S_{\nu}=\epsilon_{\nu}/\alpha_{\nu}$, where $\epsilon_{\nu}$ and $\alpha_{\nu}$ are, respectively, the line emissivity and the absorption coefficient, we can write Eq.~\ref{eq:transfer1} as
\begin{equation}
  I_{\nu}(\tau_{\nu}) \approx \frac{\epsilon_{\nu}}{\alpha_{\nu}}.
  \label{eq:transfer2}
\end{equation}

For a spectral line arising from the transition $i\rightarrow j$, the line emissivity, $\epsilon_\nu$, in erg\,s$^{-1}$\,cm$^{-3}$\,Hz$^{-1}$\,sr$^{-1}$, is given by
\begin{equation}
  \epsilon_{\nu} = \frac{1}{\Omega} \, \Delta E \, A_{ij} N_i \, \phi_\nu,
  \label{eq:epsilon}
\end{equation}
where $A_{ij}$ is the Einstein coefficient for spontaneous transition (in s$^{-1}$) between the upper level $i$ (with energy $E_i$) and the lower level $j$ (with energy $E_j$), separated by \mbox{$\Delta E = E_i - E_j$} (in erg), $N_i$ is the population of ions in the upper level (in cm$^{-3}$), and $\Omega$ is the solid angle subtended by the emitting region (in sr).

The total absorption coefficient of the transition, $\alpha_\nu$, in cm$^{-1}$, is given by
\begin{equation}
  \alpha_{\nu} = \frac{1}{\Omega} \, \Delta E \, (B_{ji} N_j - B_{ij} N_i) \, \phi_\nu,
  \label{eq:kappa}
\end{equation}
where $N_j$ is the population of ions on level $j$, $g_i$ and $g_j$ are the statistical weight of the levels, $B_{ji}$ and $B_{ij}$ are the Einstein coefficients for absorption and stimulated emission, respectively. Considering that contributions from stimulated emission are negligible relative to other line formation processes, then $B_{ij} N_i\approx 0$. Therefore, combining Eq.~\ref{eq:transfer2}, \ref{eq:epsilon}, \ref{eq:kappa}, and recalling the relations between the Einstein coefficients, the specific intensity, in erg\,s$^{-1}$\,cm$^{-2}$\,\AA$^{-1}$\,sr$^{-1}$, will be given by

\begin{equation}
  I_{\lambda} \approx \frac{2hc^2}{\lambda^5} \, \frac{g_j}{g_i} \, \frac{N_i}{N_j}.
  \label{eq:final}
\end{equation}
In order to change from sr to arcsec$^2$ (so that the specific intensity unit is erg\,s$^{-1}$\,cm$^{-2}$\,\AA$^{-1}$\,arcsec$^{-2}$), this equation must be multiplied by \mbox{$(\pi/(3600\times180))^2=2.35\times10^{-11}$}. Note that the time variable in equation \ref{eq:final} is the relative level population, $N_i/N_j$, which is set by the local electron density ($n_\rmn{e}$) and temperature ($T_\rmn{e}$) at each phase. Therefore, at a given epoch $\Phi$, the line brightness relative to epoch $\Phi_0$ will be given by
\begin{equation}
 \frac{I_{\lambda}(\Phi)}{I_{\lambda}(\Phi_0)} = \frac{\left(\frac{N_i}{N_j}\right)_{\Phi}}{\left(\frac{N_i}{N_j}\right)_{\Phi_0}},
 \label{eq:normalintensity}
\end{equation}
where $\Phi_0$ is an arbitrary epoch chosen as normalization factor. In this work, we have adopted $\Phi_0=12.516$.

\label{lastpage}

\bsp

\end{document}